\documentclass[11pt]{article}

\usepackage{float}
\floatstyle{plaintop}
\restylefloat{table}
\usepackage{tablefootnote}
\usepackage{setspace}
\usepackage{cite}
\usepackage[numbers,compress]{natbib}
\usepackage{amsmath,hyperref, wasysym}
\usepackage{latexsym}
\usepackage{amssymb}
\usepackage[margin=0.75in]{geometry}
\usepackage{wasysym}
\usepackage{fancyhdr}
\usepackage{graphicx}
\usepackage{times}
\usepackage{color,soul}
\usepackage{cancel}
\usepackage{multicol}
\usepackage{placeins}
\usepackage{authblk}
\usepackage{arydshln}
\usepackage{float}
\usepackage{dsfont}
\usepackage{caption}
\usepackage{subcaption}
\usepackage{booktabs}
\usepackage{enumerate}
\usepackage{enumitem}
\usepackage{adjustbox}
\restylefloat{table}
\usepackage{lipsum}
\usepackage[labelfont=bf]{caption}
\renewcommand{\figurename}{Figure}
\begin{document}
\title{How time and pollster history affect U.S. election forecasts under a compartmental modeling approach}
\author[1]{Ryan Branstetter}
\author[2]{Samuel Chian}
\author[1]{Joseph Cromp}
\author[2]{William L.\ He}
\author[2]{Christopher M.\ Lee}
\author[3]{Mengqi Liu}
\author[2]{Emma Mansell}
\author[4]{Manas Paranjape}
\author[4]{Thanmaya Pattanashetty}
\author[1]{Alexia Rodrigues}
\author[1,*]{Alexandria Volkening}

\affil[1]{Department of Mathematics, Purdue University, West Lafayette, IN}
\affil[2]{Department of Engineering Sciences and Applied Mathematics, Northwestern University, Evanston, IL}
\affil[3]{Elmore Family School of Electrical and Computer Engineering, Purdue University, West Lafayette, IN}
\affil[4]{Department of Computer Science, Purdue University, West Lafayette, IN}
\affil[*]{Email address for correspondence: \href{mailto:avolkening@purdue.edu}{avolkening@purdue.edu}}
\maketitle

\begin{abstract}
In the months leading up to political elections in the United States, forecasts are widespread and take on multiple forms, including projections of what party will win the popular vote, state ratings, and predictions of vote margins at the state level. It can be challenging to evaluate how accuracy changes in the lead up to Election Day or to put probabilistic forecasts into historical context. Moreover, forecasts differ between analysts, highlighting the many choices in the forecasting process. With this as motivation, here we take a more comprehensive view and begin to unpack some of the choices involved in election forecasting. Building on a prior compartmental model of election dynamics, we present the forecasts of this model across months, years, and types of race. By gathering together monthly forecasts of presidential, senatorial, and gubernatorial races from 2004--2022, we provide a larger-scale perspective and discuss how treating polling data in different ways affects forecast accuracy. We conclude with our 2024 election forecasts \cite{2024forecasts} (upcoming at the time of writing).
\end{abstract}


\textit{In memory of Daniel F.\ Linder, remembering his quiet nature and consistent kindness.}

\section{Introduction}\label{sec:intro}

From a scientific perspective and a policy one, political elections are of wide interest, and forecasting them is a task that invites many different approaches \cite{VolkeningChapter}. These approaches include building statistical models \cite{Abramowitz1988,Abramowitz2008,Erikson1996} that are informed by historical voting patterns, approval ratings, economic indicators, or other so-called ``fundamental data"\footnote{Fundamental data are data on various things that voters may use to choose a candidate \cite{Gelman}. These data include information about how the economy is doing, for example.},
as well as developing data-driven models \cite{Linzer,Chen2023,Heidemanns2020,Jackman,Wang} that rely at least partly on trial-heat polls. Some methods lead to a single static forecast, while others---particularly those that involve polling data---produce dynamic forecasts that are regularly updated. This raises questions about how forecast accuracy changes in time as Election Day nears. Moreover, because many organizations conduct polls, it is natural to ask whether accounting for the historical accuracy of various pollsters may improve forecasts. Motivated by these questions, here we present a large-scale study of forecasts in time across presidential, senatorial, and gubernatorial elections from 2004--2022. Throughout this study, our work centers on a prior mathematical model of election dynamics \cite{Volkening2020}, and we investigate how time and accounting for historical pollster performance affects its forecasts.

Journalists, media outlets, websites, blogs, and researchers generate election forecasts at various scales with various data. Forecasters may present the chance of a Democratic or Republican candidate winning the Electoral College, predictions of the national popular vote, ratings of each state race's competitiveness, or forecasts of state vote margins. In the political-science community, for example, it has been common to make national-level forecasts based on fundamental data, sometimes in combination with select polls. Such approaches \cite{Wlezien1996,Abramowitz1988,Abramowitz2008,Lewis-Beck1984,Wlezien2004} often involve regression on historical data and forecast the popular vote in presidential races. They may produce one-time, early forecasts \cite{Abramowitz1988,Abramowitz2008} or dynamic forecasts \cite{Wlezien2004,Wlezien1996}. As a different perspective, Sabato's Crystal Ball \cite{Sabato}, Inside Elections \cite{InsideElections}, and the Cook Political Report \cite{Cook,Cook2014} categorize each state with a rating such as ``safe", ``likely", or ``toss-up", using qualitative and quantitative data.

Forecasts of state-level vote margins typically involve polling data. Early examples include the work of Campbell \cite{Campbell1992}, who combined fundamental data with select national polls. Depending more on polls, RealClearPolling \cite{RealClearPolling}, which is associated with RealClearPolitics \cite{RealClearPolitics2004,RealClearPolitics2008}, aggregates state and national polls from many sources and dynamically summarizes them; their website features projections of electoral votes, state ratings, and state-level vote margins. In a related vein, Wang \cite{Wang} and his team at the Princeton Election Consortium \cite{PrincetonElectionConsortium} apply statistical methods to compute state-level median margins from recent polls and report electoral-vote distributions, among other things. There is also a large group of analysts who combine state polls with fundamental data, taking a Bayesian perspective \cite{Linzer,Chen2023}. Linzer \cite{Linzer} produced dynamic state forecasts by blending a historical model with polls, and Heidemanns et al.\ \cite{Heidemanns2020} built on this approach for \textit{The Economist}'s \cite{TheEconomist} $2020$ presidential forecasts. As another prominent mainstream source, Nate Silver \cite{SilverBook,SilverBulletin} and/or\footnote{Silver \cite{SilverBook,SilverBulletin} and FiveThirtyEight (at ABC News in recent years) have separated \cite{SilverBulletinAnnoucement}. Up until and including the 2022 elections that we discuss, the two were associated. Silver has since been posting 2024 forecasts via the \textit{Silver Bulletin} \cite{SilverBulletin}, and 2024 forecasts are under ``538" through ABC News \cite{538mainNew}.}
the team at FiveThirtyEight \cite{538mainNew,538mainOld} provide detailed forecasts---including projected state margins---on their websites.

\begin{figure}[t!]
\includegraphics[width=\textwidth]{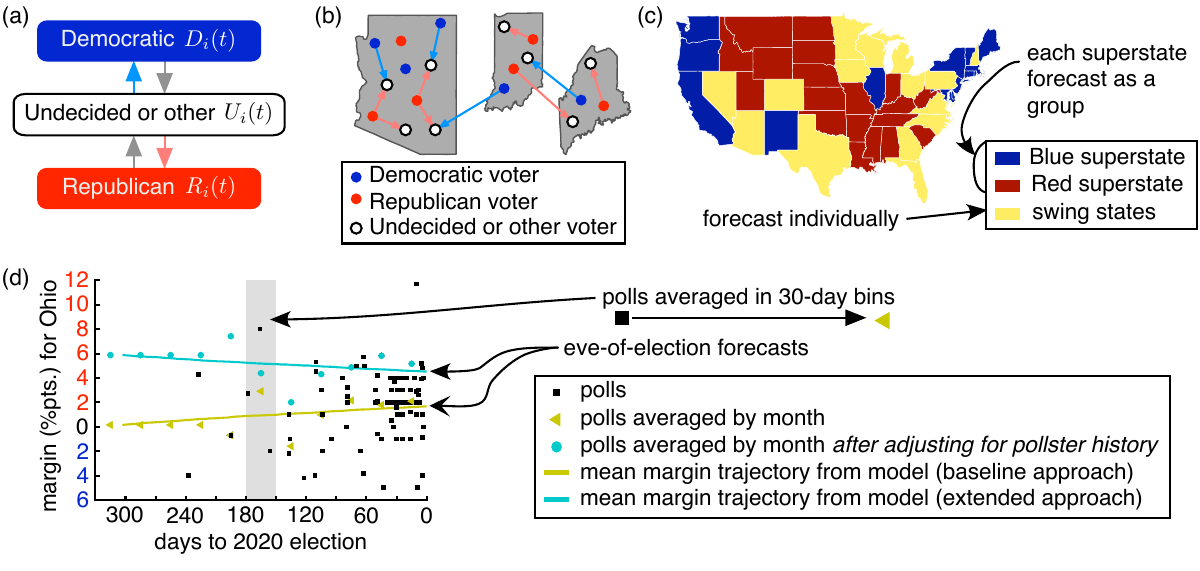}
\vspace{-1\baselineskip}
\caption{Overview of our modeling approach. (a) A compartmental Republican--Undecided--Democratic model \cite{Volkening2020} is the basis of our work. Borrowing ideas from Susceptible--Infected--Susceptible models \cite{ccc,Diekmann,HethcoteReview,Kermack700}, this approach \cite{Volkening2020} treats Democratic and Republican voting intentions as two types of ``contagion". As in \cite{Volkening2020}, we track the fractions of Democratic ($D_i(t)$), Republican ($R_i(t)$), and undecided or other voters in time in each region $i$. (b) According to the model \cite{Volkening2020}, committed voters may change to undecided (or other), and undecided voters may become Democratic or Republican through opinion ``transmission". Interactions between undecided and committed voters may occur within or between states, and they need not be symmetric. (c) As in \cite{Volkening2020}, we reduce the number of equations by combining reliably Democratic (respectively, Republican) states into a ``Blue superstate" (respectively, ``Red superstate"). For each superstate, we generate one forecast; we forecast swing states (yellow) individually. See the supplementary material for our superstate definitions. (d) Our work relies on polls, and we consider two ways of treating these data: in our ``baseline approach" \cite{Volkening2020}, we sort polls (black squares) into $30$-day bins and take the average within each bin to produce the data (green squares) that we use to fit model parameters. After selecting parameter values by minimizing the difference between the deterministic model's \cite{Volkening2020} solution and these monthly data points, we simulate the stochastic version of the model many times and take the mean result at zero days before the election as the forecast margin, as in \cite{Volkening2020}. In our ``extended approach", we adjust the polls prior to parameter fitting to account for historical differences between polling houses. We abbreviate percentage points as ``\%pts." Black points in (d) are based on polls aggregated by FiveThirtyEight \cite{FiveThirtyEightLatestPolls}. \label{fig:intro}}
\end{figure}

In terms of methods, as we highlight above, election forecasts have predominantly resulted from regression-based models, Bayesian statistical perspectives, and political-science groups. While opinion dynamics \cite{castellano09,PorterGleeson,Miller2020} are the subject of much research in applied dynamical systems, 
the mathematical-modeling community working on topics related to political opinion dynamics has focused on questions outside of forecasting (e.g., \cite{Braha,GalamTrump, Yang2020,Fernandez2014,BottcherPLOS,Biondo2018,Hernandez2018}), with few exceptions \cite{Volkening2020}. For example, Restrepo et al.\ \cite{Restrepo2009} developed a compartmental model to investigate the effects of exit polls on voter turnout, and Fern\'{a}ndez-Gracia et al.\ \cite{Fernandez2014} studied spatial correlations in election results. Instead interested in real-time forecasting, Volkening, Linder, Porter, and Rempala \cite{Volkening2020} developed a dynamical-systems perspective that relies on compartmental modeling and state-level polls. Their model takes the form of a system of differential equations tracking the fractions of undecided, Republican, and Democratic voters in different states; see Fig.~\ref{fig:intro}(a)--(c). Using polling data to determine model parameters, Volkening et al.\ \cite{Volkening2020} applied their approach to presidential, senatorial, and gubernatorial elections in $2012$, $2016$, and $2018$, with an emphasis on eve-of-election forecasts in swing states. Their model performed similarly to FiveThirtyEight \cite{538mainOld} and Sabato's Crystal Ball \cite{Sabato}, with the added strength of being fully transparent \cite{Gitlab_elections}. We built on the model \cite{Volkening2020} to offer real-time forecasts of the $2020$, $2022$, and $2024$ elections on our websites \cite{2020forecasts,2022forecasts,2024forecasts}.

\begin{figure}[t!]
\includegraphics[width=\textwidth]{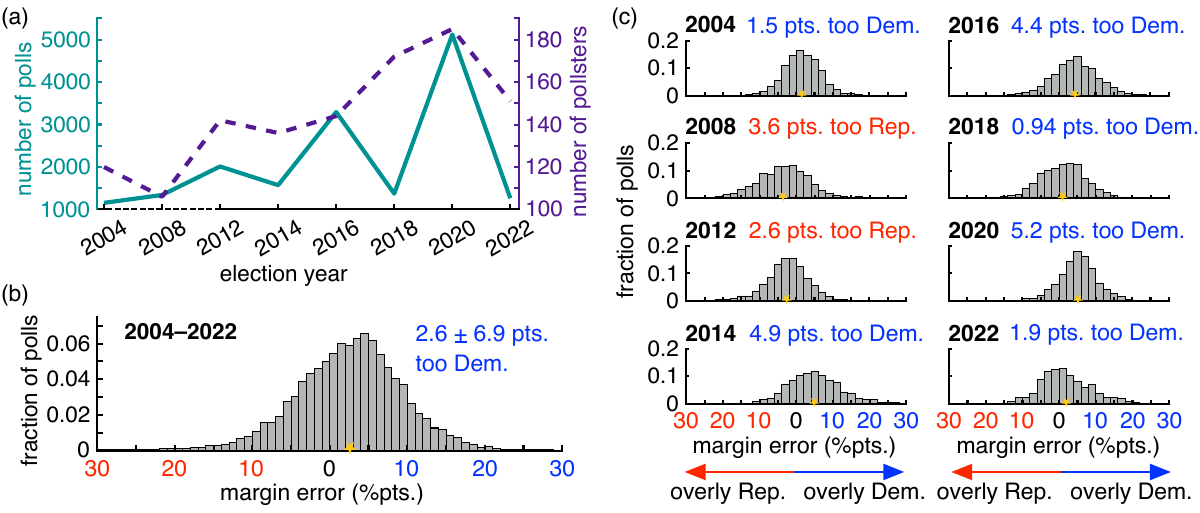}
\vspace{-\baselineskip}
\caption{Summary of the polling data that we use for forecasting. (a) The number of state-level polls tends to be much higher for presidential elections, as compared to Senate and governor races \cite{Chen2023,Shirani-Mehr2018}. We also show the number of pollsters in time that we identify as responsible for these polls; see Sect.~\ref{sec:polling}. (b) For presidential elections from $2004$--$2020$, as well as senatorial and gubernatorial elections from $2012$--$2022$, the state-level polls tend to lean Democratic. (c) Breaking down the data in panel (b) by year, we observe that the polls have shifted from leaning Republican in $2008$ and $2012$ to leaning Democratic in the last decade. Data in (a)--(c): Depending on the year, we use polling data from FiveThirtyEight \cite{FiveThirtyEightLatestPolls,FiveThirtyEightGitHub}, HuffPost Pollster \cite{HuffPostPollster,HuffPostAPI}, or RealClearPolitics \cite{RealClearPolitics2004,RealClearPolitics2008,RealClearPollingData}; we also build on the data formatted by Volkening et al. \cite{Volkening2020} for $2012$ and $2016$ \cite{Gitlab_elections}. We do not include national polls, and we limit to polls within $330$ days of Election Day.\label{fig:data}}
\end{figure}

Whether statistical \cite{Wang,538mainOld,538mainNew,Linzer} or mathematical \cite{Volkening2020}, models that rely on aggregated polls face questions about how to merge and adjust these data \cite{Pasek2015}. Polls may differ from the election results for many reasons, including random variation, errors due to non-response or misidentification of likely voters, pollster herding,
and the presence of late deciders \cite{Pasek2015,Shirani-Mehr2018,Pickup2008,AAPOR,prosser_mellon_2018,PANA2009}. Moreover, polls are conducted by polling organizations (or ``polling houses") of differing quality, and their individual methodologies can lead to ``house effects", tendencies to consistently lean Republican or Democratic \cite{deStefano2022,Gelman2021,Pasek2015,Selb2016,Traugott2014,Jackman}. Poll accuracy also depends on the time to Election Day, and there are differences in the magnitude of errors in polls of presidential, senatorial, and gubernatorial elections \cite{Shirani-Mehr2018,Campbell}. Looking across cycles, as we show in Fig.~\ref{fig:data}, the number of polls and distribution of poll errors has changed over the last twenty years, and the state-level polls have collectively leaned Democratic since $2014$.

Approaches to addressing polling errors range from simplified to complex.
One option \cite{Wang,Volkening2020} is to treat all polls the same and essentially assume that aggregating the polls will address the errors. Volkening et al. \cite{Volkening2020} took this perspective: they did not consider the historical accuracy of pollsters or house effects. Stepping into more complexity, several statistical studies have accounted for house effects in polls before aggregating \cite{Jackman,Heidemanns2020}. A more complex option is exemplified by FiveThirtyEight \cite{538mainNew,538mainOld}, the \textit{Silver Bulletin} \cite{SilverBulletin}, and \textit{The New York Times}. Silver \cite{SilverBook,SilverBulletin} and FiveThirtyEight \cite{538mainOld,538mainNew} undertake many pollster- and time-specific adjustments. This includes weighting polls using their ``pollster ratings", correcting for house effects, adjusting polls of registered voters or all adults to frame these data as polls of likely voters, and adjusting the polls around conventions. All of this raises questions about how each of these adjustments affects accuracy, and about how forecasts evolve in time.

With these questions in mind, here we take a more comprehensive, time-dynamic view, presenting monthly forecasts from July to November of elections across the last two decades. We center our work on the mathematical model  \cite{Volkening2020}, which we used to post real-time forecasts of the $2020$ and $2022$ elections on our websites \cite{2020forecasts,2022forecasts}. Unlike some work that focuses on presidential or senatorial elections and final forecasts, we present regular forecasts across multiple presidential, senatorial, and gubernatorial cycles. This allows us to investigate how forecast accuracy differs between election types and to consider two time scales (within a cycle and across cycles). In addition to our study of the model \cite{Volkening2020} in time, we also begin to consider the role of alternative choices, focusing on the impact of incorporating historical information about pollster performance into forecasts. To do so, we undertake a large study to reconcile different data sources and pollster names across the last twenty years. Our code and data are publicly available in GitLab \cite{Gitlab_elections_v2}, and we hope our work leads to more research on elections from a dynamical-systems perspective. We conclude with our forecasts of the 2024 elections, which are in the future at the time of this writing; also see our website \cite{2024forecasts}.

\section{Methods}\label{sec:methods}

\begin{figure}[t!]
\includegraphics[width=\textwidth]{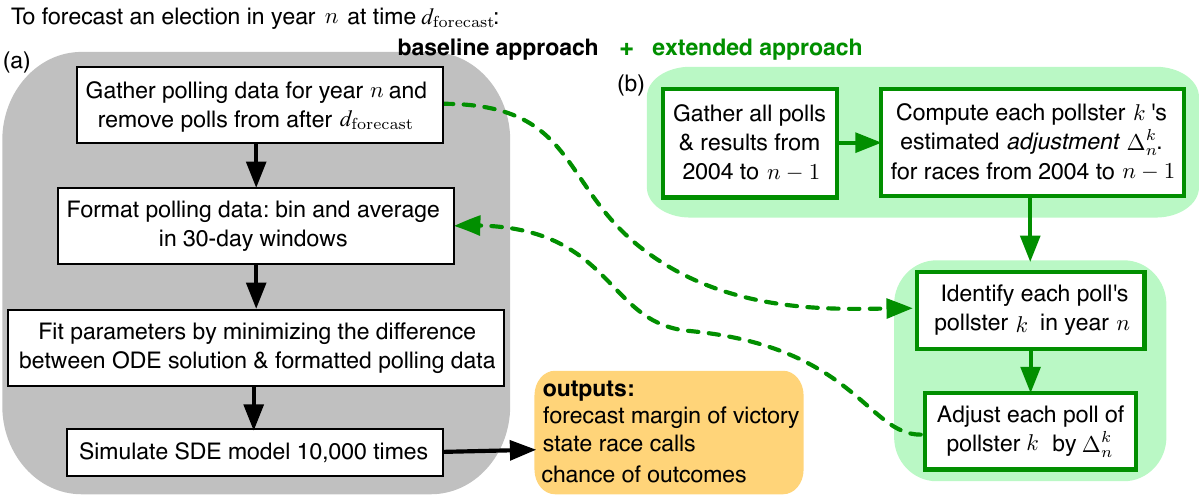}
\vspace{-\baselineskip}
\caption{Summary of our baseline and extended approaches in our forecasting pipeline. Our baseline approach follows the one in \cite{Volkening2020} developed for elections in $2012$, $2016$, and $2018$, and our extended approach accounts for historical pollster performance. (a) To forecast an election in year $n$ on a date $d_\text{forecast}$ before Election Day using our baseline approach, we gather year $n$'s polls and remove any polls with end dates after $d_\text{forecast}$; bin the remaining polls and average them in $30$-day windows by state, as we describe in Fig.~\ref{fig:intro}(d); fit model parameters by minimizing the difference between the solution to Eqns.~\eqref{eq:ode1}--\eqref{eq:ode3} and the month-averaged polling data; and simulate Eqns.~\eqref{eq:sde1}--\eqref{eq:sde2} $10,000$ times from 1 January to Election Day with these parameters \cite{Volkening2020}. We then use the simulation results on Election Day to report the forecast margin of victory (i.e., mean margin across $10,000$ simulations), race calls (i.e., Democratic or Republican outcome), and ratings (i.e., chances of outcomes). (b) Our extended approach begins by gathering all past (from the perspective of year $n$) polls and results for elections between $2004$ and year $n-1$, inclusive.  After accounting for differences in pollster naming conventions in time, we compute each pollster $k$'s ``adjustment value" $\Delta^k_n$ for use when forecasting elections in year $n$. For each poll before $d_\text{forecast}$ in year $n$, we identify the poll's pollster and adjust the poll's Democratic and Republican percentages to account for their average historical lean $\Delta^k_n$ that we estimate. We then use these adjusted polls as the data that we bin and average in $30$-day windows, fitting parameters and simulating opinion dynamics as in the baseline approach \cite{Volkening2020}. 
\label{fig:pipeline}}
\end{figure}

As we show in Fig.~\ref{fig:pipeline}, we use two forecasting approaches: a ``baseline approach" and an ``extended approach". Both approaches build on the model \cite{Volkening2020} of presidential, senatorial, and gubernatorial elections in $2012$, $2016$, and $2018$. 
Broadly, our baseline approach involves gathering polls, formatting these data, estimating model parameters, and simulating opinion dynamics. Real-time forecasts using our baseline approach are available on our websites for $2020$ \cite{2020forecasts}, $2022$ \cite{2022forecasts}, and $2024$ \cite{2024forecasts}. In our extended approach, we adjust the polls to account for historical pollster performance before estimating parameters; see Fig.~\ref{fig:intro}(d).

\subsection{Compartmental Republican--Undecided--Democratic model}\label{sec:model}

The model \cite{Volkening2020} that is the basis of our work is a compartmental model at its core. One assumption behind compartmental models is that individuals can be divided into groups, with rules determining when group members swap compartments \cite{Volkening2020,ccc}. For example, in the Susceptible--Infected--Susceptible (SIS) model \cite{Kermack700,HethcoteReview}, individuals are \textit{susceptible} or \textit{infected}. Susceptible individuals become infected by interacting with members of the infected compartment, and infected individuals recover. Beyond disease transmission, compartmental models also have a long history of being applied to social dynamics \cite{IdeaSpread,FrenchRiots,Marvel2012}. Volkening et al.'s compartmental model \cite{Volkening2020} tracks the fractions of Democratic ($D_i$), Republican ($R_i$), and undecided ($U_i$) voters within each region of interest (i.e., state) $i$ using systems of ordinary or stochastic differential equations (ODEs or SDEs). See~Fig.\ \ref{fig:intro}(a)--(b). Specifically, in the deterministic case, we specify that voter dynamics in region $i$ follow the ODEs \cite{Volkening2020}:
\begin{align}
    \frac{dR_i}{dt}(t) &= -\gamma_\text{R}^i R_i + \sum_{j=1}^M \beta_\text{R}^{ij} \frac{N^j}{N} U_i R_j, \label{eq:ode1} \\
    \frac{dU_i}{dt}(t) &= \underbrace{\gamma_\text{R}^i R_i}_\text{Rep.\ ``loss"} + \underbrace{\gamma_\text{R}^i R_i}_\text{Dem.\ ``loss"} ~\underbrace{- ~\sum_{j=1}^M \beta_\text{R}^{ij} \frac{N^j}{N} U_i R_j}_\text{Rep.\ ``transmission"}~ \underbrace{- ~\sum_{j=1}^M \beta_\text{D}^{ij} \frac{N^j}{N} U_i D_j}_\text{Dem.\ ``transmission"},\label{eq:ode2}  \\
    \frac{dD_i}{dt}(t) &= - \gamma_\text{D}^i D_i + \sum_{j=1}^M \beta_\text{D}^{ij} \frac{N^j}{N} U_i D_j; \label{eq:ode3} 
\end{align}
or, in the stochastic case, the corresponding SDEs \cite{Volkening2020}:
\begin{align}
    dD_i(t) &= \left(-\gamma_\text{D}^i D_i + \sum_{j=1}^M \beta_\text{D}^{ij} \frac{N^j}{N} U_i D_j\right)dt + \sigma d{W_\text{D}^i}(t), \label{eq:sde1} 
    \\
    dR_i(t) &= - \gamma_\text{R}^i R_i + \sum_{j=1}^M \beta_\text{R}^{ij} \frac{N^j}{N} U_i R_j + \sigma d{W_\text{R}^i}(t), \label{eq:sde2} 
\end{align}
where we note that $U_i = 1 - R_i - D_i$. Compartmental models are often abbreviated using the first letters of their compartments, so we call Eqns.~\eqref{eq:ode1}--\eqref{eq:sde2} the ``cRUD" model \cite{2022forecasts,2024forecasts}.

The parameters $\gamma_\text{D}^i$ and $\gamma_\text{R}^i$ in Eqns.~\eqref{eq:ode1}--\eqref{eq:sde2} are the Democratic and Republican ``recovery" rates of undecided voters in region $i$. Similarly, the parameters $\beta_D^{ij}$ and $\beta_R^{ij}$, respectively, are the Democratic and Republican ``transmission" rates\footnote{As in \cite{Volkening2020}, we stress that opinion dynamics and disease transmission are certainly different. Using epidemiological language simply serves to show that the models in Eqns.~\eqref{eq:ode1}--\eqref{eq:sde2} are compartmental models.} from region $j$ to region $i$. We determine the values of these parameters for an election in year $y$ using year $y$'s publicly available polling data, as we describe in Sect.~\ref{sec:parameters}. This means that our parameter values differ across years and across presidential, senatorial, and gubernatorial elections. The year-specific values of $N^j$, the number of voting-age people in region $j$, and $N$, the number of voting-age individuals in the country, are based on data from the Federal Register \cite{FedRegister2004,FedRegister2008,FedRegister2012,FedRegister2014,FedRegister2016,FedRegister20202018,FedRegister2022}. In the stochastic version of the model \cite{Volkening2020}, $W_\text{D}^i$ and $W_\text{R}^i$ are Wiener processes, and $\sigma = 0.0015$ is the noise strength. Following the approach in \cite{Volkening2020}, we correlate noise in Eqns.~\eqref{eq:sde1}--\eqref{eq:sde2} based on demographic data \cite{DemographicData2020201820162014201220082004,EducationData2020201820162014201220082004,DemographicData2022,EducationData2022}; see Appendix~\ref{sec:appJaccard} for details.

The authors \cite{Volkening2020} provide forecasts in $M$ regions, and, in most cases $i \in \{1,2,...,M\}$ refers to a specific swing state. However, to reduce the number of parameters and account for limited polling of non-competitive states, Volkening et al. \cite{Volkening2020} forecast safe Democratic and Republican races as two collectives by introducing a ``Blue superstate" and ``Red superstate". For each superstate, the authors \cite{Volkening2020} produce one conglomerate margin of victory based on all of their combined polls. This means that the number $M$ of differential equations for $R(t)$ (or $D(t)$) is typically much less than the total number $\tilde{M}$ of state or district races taking place in a given cycle. The states and districts\footnote{Presidential elections in the U.S. take place in $50$ states and one district, the District of Columbia.} that we forecast as part of our superstates depend on the year, and they differ between presidential, senatorial, and gubernatorial elections. See Fig.~\ref{fig:intro}(c) for the superstate definitions that we use for the $2020$ presidential elections and the supplementary material for a complete list of superstate definitions.

\subsection{Polling data and parameter estimation} \label{sec:parametersAll}

As we show in Fig.~\ref{fig:intro}(d), our baseline (Sect.~\ref{sec:parameters}) and extended approaches (Sect.~\ref{sec:polling}) differ in terms of how we adjust and format polls to prepare these data for parameter estimation.

\subsubsection{Our baseline approach}\label{sec:parameters}

In our baseline approach, we follow conceptually the same methods \cite{Volkening2020} to format polling data and estimate parameters in Eqns.~\eqref{eq:ode1}--\eqref{eq:sde2}, with a few updates to make them more amenable to forecasting in the months before Election Day and to account for polling data aggregated by different sources. As we show in Figures~\ref{fig:intro}(d) and \ref{fig:pipeline}, this process involves gathering the polls for the election (presidential, senatorial, or gubernatorial) of interest in year $y$, sorting and averaging these data in $30$-day bins for each state or superstate, and selecting model parameters by minimizing the difference between these formatted data and the deterministic solutions of Eqns.~\eqref{eq:ode1}--\eqref{eq:ode3}.

To forecast an election in year $y$ on date $d_\text{forecast}$, we begin by downloading the associated polls and removing any with end dates strictly after $d_\text{forecast}$. We use publicly available polls aggregated by FiveThirtyEight \cite{FiveThirtyEightLatestPolls,FiveThirtyEightGitHub} (for elections in $2024$, $2022$, $2020$, and $2018$), HuffPost Pollster \cite{HuffPostPollster,HuffPostAPI} (for $2016$, $2014$, and $2012$), and RealClearPolitics \cite{RealClearPolitics2004,RealClearPolitics2008,RealClearPolling} (for $2008$ and $2004$). Our next step is cleaning the data. For instance, we do not include national polls or polls for states holding elections on dates\footnote{As an example, Georgia held a run-off election on $5$ January $2021$ following the $2020$ senatorial elections, so we filter out polls pertaining to the $5$ January election. We also do not consider polls specifically associated with the run-off election for the 2018 Senate race in Mississippi.} other than Election Day. For elections from $2018$ onward, our data may include polls for congressional districts in Maine and Nebraska; we remove any such polls and focus on polls at the state level. Because our polls are from FiveThirtyEight \cite{FiveThirtyEightLatestPolls,538mainOld,538mainNew} for $2018$--$2024$, these data include candidate names and parties. In these cases, we remove any polls that are not for the main Democratic and Republican candidates\footnote{
Because of Louisiana's special voting system, we filter out Louisiana's Senate polls in 2020 and 2022. We also do not include Alaska's 2022 polls because it used rank-choice voting. (For these elections, we treat Louisiana and Alaska as part of the Red superstate.) In the 2018 senatorial elections, we consider the Independent candidates in Vermont and Maine to be Democratic. The 2020 Arkansas Senate race featured a Republican candidate versus a Libertarian candidate (Harrington); we treat Harrington as a Democrat for this race. For the 2020 Senate election in Utah, we consider McMullin (an Independent) the Democrat running for the purposes of our model. Similarly, we consider the Independent candidates running in Vermont, Maine, and Nebraska to be Democrats in our 2024 forecasts. In terms of identifying the main Republican and Democrat running, this is not always clear until August or September for senatorial and gubernatorial elections. When we present early forecasts in this study, we use the eventual candidates as our main Republican and Democratic candidates.}. 

Once the data are clean, we format them in preparation for parameter estimation, as in \cite{Volkening2020}. For each state or district, we sort its polls into eleven $30$-day bins based on the time between their middle dates (the mean of each poll's start and end dates) and Election Day. These $30$-day bins represent months under the simplifying assumption that each month is $30$ days long. For example, the earliest bin (our ``January bin") holds polls that are $T \in [330,300)$ days before Election Day, the next earliest bin (our ``February bin") holds polls that are $T \in [300,270)$ days before Election Day, and the last bin (our ``November bin") holds polls that are $T \in [30, 0]$ days before Election Day. Because we remove polls with end dates after $d_\text{forecast}$ prior to binning, some of the latter bins will be empty when we produce early forecasts. We define $T_\text{forecast}$ to be the number of $30$-day bins from $330$ days before Election Day to $d_\text{forecast}$, where we allow the last bin to potentially be less than $30$ days long. We only consider these $T_\text{forecast}$ bins when fitting parameters to generate a forecast on $d_\text{forecast}$.

For each state, within each non-empty bin, we find the state's mean Democratic fraction and mean Republican fraction across the polls in that bin. As in \cite{Volkening2020}, if a state has no polls in some---but not all---of its $T_\text{forecast}$ bins, we define the bin-averaged data point for that state using linear interpolation. If the bin(s) with missing data occurs early in the year before the state has any polls, we find the earliest bin with polls and assign its average to all of the months before it. While more rare, it is possible for a state to have no polls near Election Day; in this case, we find the most recent bin with data and set the remainder of the $T_\text{forecast}$ bins for that state to it average. For any superstates, we follow this by computing a weighted average across the mean Democratic and Republican fractions of states with some polling data in each superstate, weighting by the state voting-age populations \cite{FedRegister2004,FedRegister2008,FedRegister2012,FedRegister2014,FedRegister2016,FedRegister20202018,FedRegister2022}.

Because our study investigates the role of time in forecast accuracy, it is common for some states to have no polls in our data set for some choices of $d_\text{forecast}$. If a state has all empty bins, indicating that it has no polls on or before $d_\text{forecast}$ and within $330$ days of the election, this presents a challenge, as we rely on polls for parameter estimation. If the state is a swing state, we reduce the number of regions in our model by one. For completeness when presenting and evaluating forecasts in Sect.~\ref{sec:results}, we then interpret this state as having a forecast margin of victory of zero and a $50$--$50$\% chance of leading to a Republican or Democratic outcome. Alternatively, if the state without polls is part of a superstate, this presents no issue: as in \cite{Volkening2020}, such a state receives the superstate's forecast as its forecast.

At this stage, we have formatted data $\{\tilde{D}_i(T_j), \tilde{R}_i(T_j)\}_{i=1,...,M;j=1,...,T_\text{forecast}}$, where $M$ is the number of swing states or superstates, and $\tilde{D}_i(T_j)$ is the mean Democratic fraction for state or superstate $i$ in month (e.g., $30$-day bin) $j$. Similarly $\tilde{R}_i(T_j)$ is the mean Republican fraction that we find for region $i$ in month $j$ with our binning procedure. We select the parameters in Eqns.~\eqref{eq:ode1}--\eqref{eq:sde2} using these data $\{\tilde{D}_i(T_j), \tilde{R}_i(T_j)\}_{i=1,...,M;j=1,...,T_\text{forecast}}$ and the deterministic version of the cRUD model. Specifically, we follow the approach \cite{Volkening2020}: we first define the vector holding our formatted data points for month $T_j$, namely:
\begin{align*}
    \tilde{\textbf{C}}(T_j) &= [\tilde{R}_1(T_j), \tilde{R}_2(T_j),...,\tilde{R}_M(T_j),\tilde{U}_1(T_j), ...,\tilde{U}_M(T_j),\tilde{D}_1(T_j), ...,\tilde{D}_M(T_j)],
\end{align*}
where $\tilde{U}_i(T_j) = 1 - \tilde{D}_i(T_j) -\tilde{R}_i(T_j)$. We then fit the parameters $\{\beta,\gamma \}$ in Eqns.~\eqref{eq:ode1}--\eqref{eq:sde2} by minimizing the least-squares difference between the data $\tilde{\textbf{C}}$ and the deterministic solutions $\textbf{C}$ to Eqns.~\eqref{eq:ode1}--\eqref{eq:ode3} across time, as below:
\begin{align}
    \{\beta,\gamma\} &= \text{argmin}_{\{\hat{\beta},\hat{\gamma}\}} \sum_{j=1}^{T_\text{forecast}} \| \tilde{\textbf{C}}(T_j) - \textbf{C}(T_j;\hat{\beta},\hat{\gamma}) \|^2_2,\label{eq:argmin}
\end{align}
where 
\begin{align*}
    \textbf{C}(T_j;\hat{\beta},\hat{\gamma}) &= [R_1(t_{T_j}), R_2(t_{T_j}),...,R_M(t_{T_j}),U_1(t_{T_j}), ...,U_M(t_{T_j}),D_1(t_{T_j}), ...,D_M(t_{T_j})]
\end{align*}
is the solution to Eqns.~\eqref{eq:ode1}--\eqref{eq:ode3} at the time $t_{T_j}$ associated with month $T_j$ under the parameters $\{\hat{\beta},\hat{\gamma}\}$. In this process, we consider the solutions of Eqns.~\eqref{eq:ode1}--\eqref{eq:ode3} over $30 \times T_\text{forecast}$ days, approximately the time between $1$ January of year $y$ and $d_\text{forecast}$. See Appendix~\ref{sec:appNumerical} for numerical-implementation details. We use these same parameter values $\{\beta,\gamma\}$ in the stochastic version of the cRUD model, Eqns.~\eqref{eq:sde1}--\eqref{eq:sde2}.

The original study \cite{Volkening2020} focuses on eve-of-election forecasts for $2012$ and $2016$, and presents forecasts in time for the $2018$ senatorial and gubernatorial elections. In these two cases, the authors \cite{Volkening2020} remove polls based on the middle date between a poll's start and end date. As we show in Fig.~\ref{fig:pipeline}, we remove polls with end dates after the forecast date $d_\text{forecast}$. The original software \cite{Gitlab_elections} includes special cases with instructions on what states have polls for given allowable forecast dates. Investigating the role of time in forecasting calls for a more adaptive approach. Similarly, because we consider a wider range of elections from $2004$--$2024$, it is necessary to use aggregated polls from more sources. Building on Volkening et al.'s code \cite{Gitlab_elections,Volkening2020}, we provide more flexible software for formatting data that adapts to changes in the number of polls available in time on GitLab \cite{Gitlab_elections_v2}.

\subsubsection{Our extended approach}\label{sec:polling}

As we discuss above, our baseline approach \cite{Volkening2020} relies on averaging each state's polls by month to produce data points for use in parameter estimation. This choice to simply average the polls---without considering house effects---mirrors the work of Wang \cite{Wang} and the Princeton Election Consortium \cite{PrincetonElectionConsortium}. On the other hand, Silver \cite{SilverBulletin}, FiveThirtyEight \cite{538mainOld,538mainNew}, and the \textit{The New York Times} \cite{NYT} adjust polls in many ways, including to account for house effects or tendencies. With this in mind, our extended approach involves adjusting the polling data---prior to fitting parameters---to incorporate the historical tendencies of pollsters\footnote{We specifically estimate the mean signed error in vote margin for each polling entity.}; see Fig.~\ref{fig:intro}(c). Our methods for estimating the historical tendencies of pollsters have important simplifications, and we stress these throughout this manuscript.

Detecting house effects in polls is challenging, and we highlight four perspectives on doing this. First, it is appealing to define house effects using the mechanisms that we think are responsible for them \cite{Pickup2008}. However, the specific choices about methodology that pollsters make are often proprietary \cite{Pickup2008}. As a second option, Jackman \cite{Jackman} used the election results to back out the house effects active in polls for that election, but this does not make sense for real-time forecasting. A third option is to assume that the net partisan lean across all of the polls is zero or some other value for the election in question \cite{Pickup2008,Heidemanns2020,Shirani-Mehr2018}, and then use this to anchor estimates of poll bias from a Bayesian perspective. This approach can suffer if industry bias causes all of the polls to collectively miss the result \cite{Pickup2008}. It is also unclear how it relates to pollster herding. A fourth option, which Jackman \cite{Jackman} also suggests, is to estimate house effects based on how pollsters performed in past races \cite{Selb2016}. Option 4 works for real-time forecasting, but it assumes pollster history is indicative of pollster future. This is problematic because pollster methods may change between elections \cite{Selb2016,Selb2023}. Moreover, reconciling the names of pollsters across decades and data sources presents a major data challenge. Despite these drawbacks, we choose this approach because we view estimating pollster tendencies based on historical data as opening many directions for future work (see Sect.~\ref{sec:conclusions}).

In order to estimate the historical tendency of each each polling organization, we must identify the pollster responsible for each poll.
While seemingly straightforward, this presents a significant hurdle. Across the polls in our data set from $2004$--$2022$, there are over $1,500$ different strings denoting pollster names. In some cases, two names are different due to a simple fix, like the presence of additional spaces or because we did not copy the pollster's full name in the case of the 2004 and 2008 data, which we gathered from RealClearPolitics \cite{RealClearPolitics2004,RealClearPolitics2008}. In other cases, the name of the organization that conducted the poll is abbreviated or ambiguous: for example, ``American Research Group" occasionally shows up as ``ARG", and ``Angus Reid Global" is written as ``Angus-Reid". Another issue is pollsters with similar names: for example, based on \cite{Bigten}, it appears that ``Big 10 Battleground" polls and ``Big Ten" polls are by the same organization. However, ``Bluegrass Data" and ``Bluegrass Voters Coalition" appear to be different. Moreover, because we are working with data from the last twenty years, some organizations have changed their name during that time frame. 

Another complication is that many polls are sponsored by media organization \textit{A} and conducted by organization \textit{B} \cite{FTEratingsWork}. Situations like these show up in our data with a ``/", as in ``Washington Post/Survey Monkey". However, a poll listed as due to ``\textit{A}/\textit{B}" can also indicate that the poll was conducted jointly by \textit{A} and \textit{B}, as FiveThirtyEight points out is the case with ``The New York Times/Siena College" polls \cite{FTEratingsWork}. In still other cases, ``\textit{A}/\textit{B}" indicates that two pollsters---neither of which is a media organization---collaborated on a poll. Because ``/" is used in many different ways, it is often unclear which organization is responsible for the polling methodology and any associated house tendencies. FiveThirtyEight \cite{FTEratingsWork} also acknowledges that there is no clear rule, and it depends on the organizations whether polls listed as due to ``\textit{A}/\textit{B}" are assigned to \textit{A}, \textit{B}, or a new polling entity ``\textit{A}/\textit{B}".

Our process for addressing these challenges relies on a combination of FiveThirtyEight's list of pollster ratings \cite{FiveThirtyEightRatings}, tracking down original websites or news coverage announcing a poll, searching for information using WayBack Machine \cite{Wayback}, cross-referencing between how polls are listed by different poll-aggregating sources\footnote{For example, if RealClearPolitics \cite{RealClearPolling} lists a poll under pollster \textit{A} with the same margin and dates as a poll listed under pollster \textit{B} by FiveThirtyEight \cite{538mainOld,538mainNew}, this supports the idea that \textit{A} and \textit{B} are the same.}, and---when the situation is still unclear---relying on a judgment call. Through this combined process, we create a library of polling-organization names, and for each name, provide a list of aliases that we judge to be referring to that same polling entity. When a pollster's name is available in FiveThirtyEight's list of pollster ratings \cite{FiveThirtyEightRatings}, we typically assign that name to be the pollster's main name in our library. The number of associated names that we identify ranges from zero to over $55$ aliases. The latter example refers to ``Public Policy Polling" (PPP), which appears in our data set gathered from HuffPost Pollster \cite{HuffPostPollster,HuffPostAPI} with many different parenthetical extensions (i.e., ``PPP (D-Vote Vets Action Fund)") to acknowledge various organizations for whom PPP conducted polls.

Across our polling data\footnote{This refers to the state-level polls taking place within $330$ days of the appropriate election day, for presidential elections from $2004$--$2020$, and for senatorial and gubernatorial elections from $2012$--$2022$. We downloaded these data from FiveThirtyEight \cite{FiveThirtyEightLatestPolls} or HuffPost Pollster \cite{HuffPostAPI,HuffPostPollster} for $2012$--$2022$, and gathered them by hand from RealClearPolitics \cite{RealClearPolling,RealClearPollingData} for $2004$--$2008$.}
 from $2004$--$2022$, we estimate that there are $481$ polling organizations; see Sect.~\ref{sec:extended2} for these results. With this library of pollster names in hand, we can now associate a pollster with each poll and estimate each pollster's historical tendencies. In this first study, we do this in the simplest way possible, computing the mean signed margin of victory error for each pollster. Thus, by ``pollster historical tendencies", we mean ``average signed pollster errors in historical elections". One weakness of this approach is that we do not distinguish between error due to sample variance and error due to house effects; see Sect.~\ref{sec:conclusions}.

Because our focus is forecasting, it is important to only consider a pollster's performance on elections prior to the race in question when estimating their historical lean. We define the estimated adjustment $\Delta_y^k$ of polling entity $k$ for use when forecasting elections in year $y$ as:
\begin{align}
    \Delta_y^k &= \frac{1}{N^k_{y-1}}\sum_{\ell=1}^{N_{y-1}^k} \left(\left(R^\text{result}_\ell - D^\text{result}_\ell \right) - \left(R^\text{poll}_\ell - D^\text{poll}_\ell \right)\right),\label{delta}
\end{align}
where $N^k_{y-1}$ is the number of polls that we estimate pollster $k$ has in our data set for elections taking place from $2004$ to $y-1$, inclusive of the endpoints; $R^\text{poll}_\ell - D^\text{poll}_\ell$ is the signed margin of victory that poll $\ell$ predicts; and $R^\text{result}_\ell - D^\text{result}_\ell$ is the true signed margin for the state or district that poll $\ell$ considers. A negative value of $\Delta_y^k$ means that we expect pollster $k$ to lean Republican in year $y$, based on their historical performance from $2004$ to $y-1$.

Our extended approach to forecasting elections in year $y$ involves using the pollster adjustments $\{\Delta_{y-1}^k\}$ to shift each poll's margin before fitting parameters; see Figures~\ref{fig:intro}(d) and \ref{fig:pipeline}. Specifically, for each poll $\ell$ by pollster $k$, we define:
\begin{align*}
\hat{R}^\text{poll}_\ell = R^\text{poll}_\ell + \frac{1}{2}\Delta_{y-1}^k ~~~~~\text{and}~~~~~
\hat{D}^\text{poll}_\ell = D^\text{poll}_\ell - \frac{1}{2}\Delta_{y-1}^k.
\end{align*}
We then bin these adjusted polls by month, compute averages to produce $T_\text{forecast}$ data points, and estimate parameters according to Eqn.~\eqref{eq:argmin} with the original averaged polling data points $\{\tilde{R}^\text{poll},\tilde{D}^\text{poll}\}$ replaced by our adjusted ones $\{\hat{R}^\text{poll},\hat{D}^\text{poll}\}$.

As an illustrative example, we take a closer look at how our extended approach affects the monthly data points $\tilde{\textbf{C}}$ that we use for parameter fitting in a few cases. When we generate our October forecast of the $2012$ presidential elections using our extended approach, our pollster adjustments result in the monthly data points for Colorado shifting by about $1.3$~\%pts.\ more Democratic on average, as compared to the monthly data points that we find using our baseline approach. Because different, though often overlapping, sets of pollsters are active in different states, the monthly data points for Florida shift by about $2.0$ \%pts.\ toward the Democratic candidate on average when we apply our extended approach in Fig.~\ref{fig:pipeline} to produce an October $2012$ forecast. On the other hand, Florida's monthly data points shift Republican by about $0.042$ and $1.3$ \%pts.\ for our extended $2016$ and $2020$ October forecasts, respectively. These shifts in the data that we use for parameter fitting in Eqn.~\eqref{eq:argmin} later lead to shifts in our forecasts, as we discuss in Sect.~\ref{sec:extended2}.

\subsection{Generating forecasts}\label{sec:summary}

After estimating parameters in the ODE version of the cRUD model for a given election based on our monthly-averaged data points of polls (baseline approach) or adjusted polls (extended approach), the last step in our forecasting pipeline is simulating the stochastic cRUD model $10,000$ times under our estimated parameter values. Specifically, we simulate Eqns.~\eqref{eq:sde1}--\eqref{eq:sde2} from $1$ January of year $y$ until Election Day, assuming each month is $30$ days long as in \cite{Volkening2020}. This means, for example, that we simulate Eqns.~\eqref{eq:sde1}--\eqref{eq:sde2} for $308$ days for elections in $2022$, because Election Day was $8$ November. See Fig.~\ref{fig:intro}(d). We interpret the mean results on Election Day (e.g., the final simulation time) across our $10,000$ stochastic simulations to be our forecast margin of victory, and we use all of the simulated results on Election Day to provide ratings of state competitiveness.

Throughout Sect.~\ref{sec:results}, we use our methods to present forecasts at monthly intervals. Our latest forecast uses $T_\text{forecast} = 11$ bins of polling data for parameter estimation and relies on polls up until and including the day before Election Day; see Sect.~\ref{sec:parameters} and Fig.~\ref{fig:intro}(d). For instance, because the $2022$ elections took place on $8$ November, our latest forecast is $7$ November. We denote this our ``November forecast". Our next latest forecast---our ``October forecast"---uses $T_\text{forecast} = 10$ bins of data and relies on polls that take place up until (and including) $31$~days before Election Day. In the case of $2022$, this corresponds to a forecast on $10$ October. More generally, when we denote a result as associated with month $T_\text{forecast}$, it can be interpreted as what our forecast would have been $30 \times (11 - T_\text{forecast}) + 1$ days before Election Day.

\subsection{Measuring forecast accuracy}\label{sec:measurement}

As stochastic, relatively infrequent events, judging election forecasts is challenging \cite{Jackman2014, prosser_mellon_2018}.
We consider three ways to quantitatively summarize forecast performance: the percentage of states in which we identify or ``call" the winning party, the margin of victory (MOV) error in swing states, and the
Brier score \cite{Brier}. Success rate is perhaps the most interesting during real-time forecasting yet the crudest in capturing the behavior of probabilistic forecasts. Specifically, as in \cite{Volkening2020}, we compute the success rate as:
\begin{align}
\text{success rate} &= 100 \times \frac{\text{number of state or district races that we call correctly}}{\text{total number of state or district races that we forecast}},\label{eq:success}
\end{align}
where this measure ``unpacks" superstates. This means that we count each state or district race that we correctly call as part of a superstate as an individual, incrementing the numerator by one. The denominator of Eqn.~\eqref{eq:success} is typically $\tilde{M}$, the total number of state- or district-level races taking place on Election Day. However, because the approach \cite{Volkening2020} assumes each election treats a Republican versus a Democrat (or, in a few cases, an Independent, Progressive, or Libertarian), we leave out states with single-party elections when computing success rate\footnote{The $2016$ and $2018$ Senate elections in California featured two Democratic candidates against one another.}. When we cannot forecast a race using our methods due to complete lack of polls\footnote{For example, there are no polls in our data set for the $2012$ gubernatorial election in Delaware. }, we interpret it as a $50$--$50$\% toss-up and a failure in terms of correctly identifying the winning party.

Our next approach to quantifying performance is the MOV error in swing states. We consider this summary statistic more telling of forecast quality than is success rate. For example, in the $2020$ presidential election, Biden, the Democratic candidate, won Arizona's electoral votes by about $0.3$ percentage points (\%pts.) over Trump, the Republican candidate. Suppose forecaster \textit{A} predicted a $20$~\%pt.-victory for Biden and forecaster \textit{B} predicted a MOV of $0.1$ \%pts. for Trump. According to success rate, \textit{A} performed better, but MOV error highlights that \textit{B} was more aware of election dynamics. Specifically, we consider:
\begin{align}\label{eq:MOV}
    \text{MOV error in swing state $i$} &= |(R_i^\text{result} - D_i^\text{result}) - (R_i^\text{forecast} - D_i^\text{forecast})|,
\end{align}
where $R_i^\text{result} - D_i^\text{result}$ is the true signed MOV, and $R_i^\text{forecast} - D_i^\text{forecast}$ is the forecast signed MOV. In the case of our forecasts, this is the mean difference between $R_i(t)$ and $D_i(t)$ in Eqns.~\eqref{eq:sde1}--\eqref{eq:sde2} at the final simulation time over $10,000$ simulations, as we discuss in Sect.~\ref{sec:summary}. When we use the mean MOV error as a summary statistic, we compute the average of the MOV errors only across swing states, the states that we forecast individually. To evaluate our forecasts, we use data on the election outcomes from \textit{Dave Leip's Atlas of U.S.\ Presidential Elections} \cite{Leip}.

Our third method for judging probabilistic forecasts is the Brier score. Wang \cite{Wang} at the Princeton Election Consortium \cite{PrincetonElectionConsortium} has used the Brier score \cite{Brier} to quantify forecast performance, and FiveThirtyEight \cite{FiveThirtyEightBrier} has discussed the Brier skill score, which treats the Brier score in comparison to some reference score. We compute the Brier score across all of the state- and district-level races that we forecast for a given election as:
\begin{align}
\text{Brier score} &= (1/\tilde{M})\sum_{i=1}^{\tilde{M}} \left(p^\text{R}_i - o^\text{R}_i\right)^2,
\end{align}
where $p^\text{R}_i$ is the Republican candidate's chance of winning state or district $i$'s race, and the indicator variable $o_i^\text{R}$ is $1$ if the Republican candidate does win and $0$ otherwise. As with success rate, when computing the Brier score for any of our forecasts, we assign each state or district within a superstate the $p^R$ value that we generate for that superstate. Each state maintains its own $o_i^\text{R}$ value. Because we assign states that we cannot forecast due to lack of polls a $50$\% chance of voting Republican, we include these states in our Brier-score computations.

Related to forecast accuracy, we comment here on how we treat run-off elections. Run-off elections are typical in states that use the so-called ``jungle-primary" or majority-vote system with many candidates on Election Day, as is the case in Louisiana. (Alaska uses a somewhat related system in some of its more recent senatorial elections.) When an election leads to a run-off election, we use the final result of the run-off election to determine the true margin of victory and winning party in all cases with one exception. The exception is the $2022$ Georgia senatorial election; in this case only, we use the original margin of victory and winning party (which ended up being the same, albeit by a larger margin, in the run-off election).

\section{Results}\label{sec:results}

For the U.S.\ presidential and midterm\footnote{For $2004$ and $2008$, we only forecast the presidential elections. Our senatorial and gubernatorial forecasts begin with the $2012$ elections, and we do not forecast senatorial and gubernatorial in off-years (years in which there are neither presidential nor midterm elections). We do not forecast House races.} elections from $2004$--$2022$, we now undertake a comprehensive study of our baseline and extended approaches. We focus on presidential elections in the main text, and present most of our senatorial and gubernatorial results in the supplementary material. To understand the role of time in forecasting, we generate forecasts at monthly intervals from July to November for each race. When quantifying our forecast accuracy, we include the corresponding forecasts from FiveThirtyEight \cite{538mainOld,538mainNew} in recent elections. We do this because FiveThirtyEight---as a popular forecaster that specifies numbers for forecast margins of victory and ratings---provides a point of reference\footnote{When we include forecasts from FiveThirtyEight \cite{538mainNew,538mainOld} in our figures, we present their forecasts on the same dates as ours if available. However, during real-time forecasting, a forecast on date $d_\text{forecast}$ does not always correspond with using polls up until and including $d_\text{forecast}$. This is partly because the time between when a poll is completed and when it appears in public data sets is not always clear. Equating $d_\text{forecast}$ with the last date on which we include polls also assumes that producing a forecast takes less than one day to advance from download-of-polling-data to announcement-of-forecast. Our methods generally take less than eight hours, depending on the computer. Because we denote FiveThirtyEight's forecasts \cite{538mainNew,538mainOld} with the date of upload, it is possible that there may be a day or so lapse between their and our forecast dates.}.

\subsection{A study of how forecast accuracy varies with time and election}\label{sec:baseline}

\begin{figure}[t!]
\includegraphics[width=\textwidth]{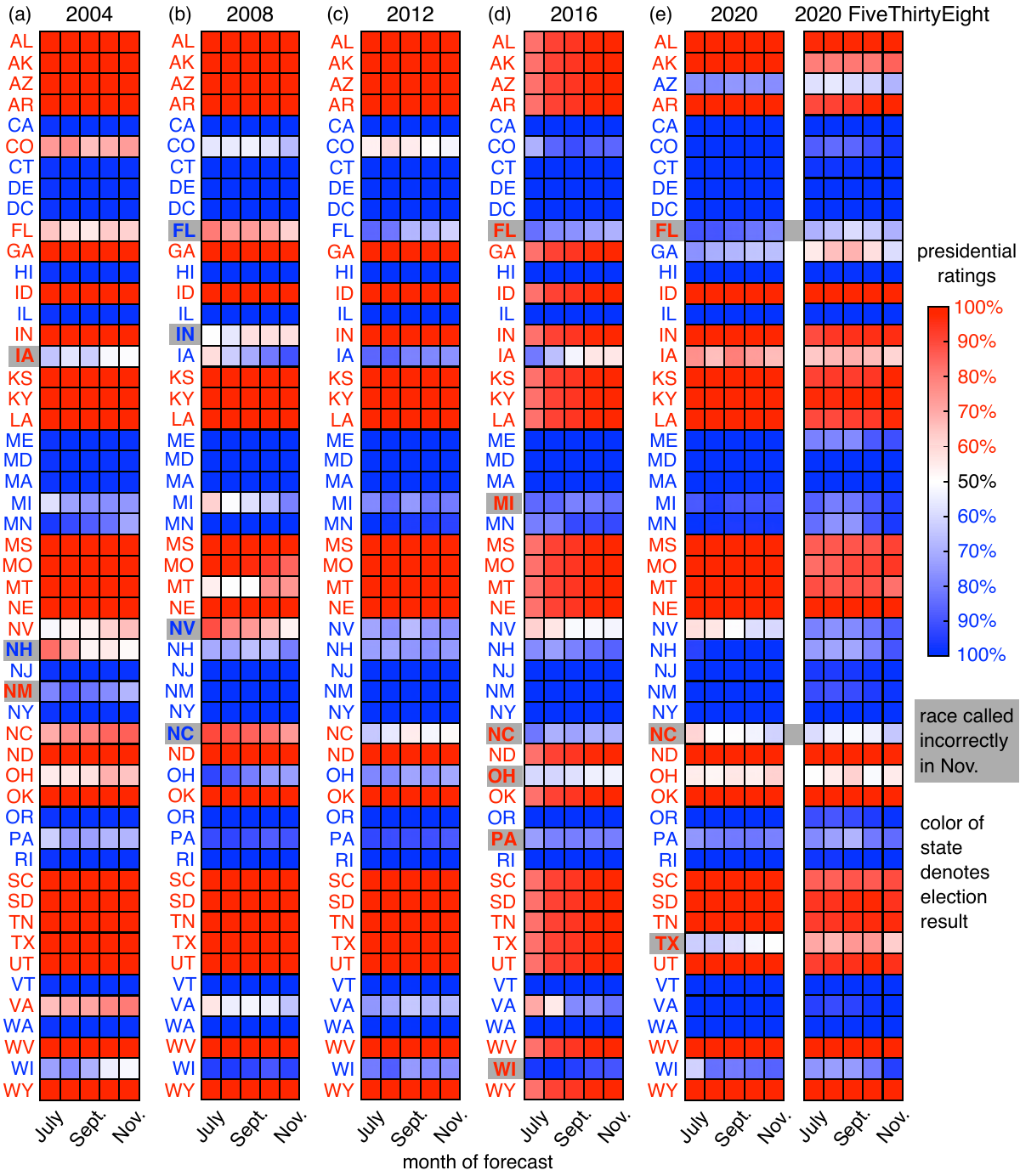}
\vspace{-1.5\baselineskip}
\caption{Ratings in time using our baseline approach \cite{Volkening2020} to forecast two decades of presidential elections. We show ratings for the races between (a) Bush (R.\, winner) and Kerry (D.); (b) McCain (R.) and Obama (D., winner); (c) Romney (R.) and Obama (D., winner); (d) Trump (R., winner) and Biden (D.); and (e) Trump (R.) and Biden (D., winner). For context, FiveThirtyEight's ratings \cite{FiveThirtyEight2020} are in the second column of (e). See Figures SM1 and SM3 in the supplementary material for corresponding senatorial and gubernatorial results.
\label{fig:forecastChancesP}}
\end{figure}

\begin{figure}[t!]
\includegraphics[width=\textwidth]{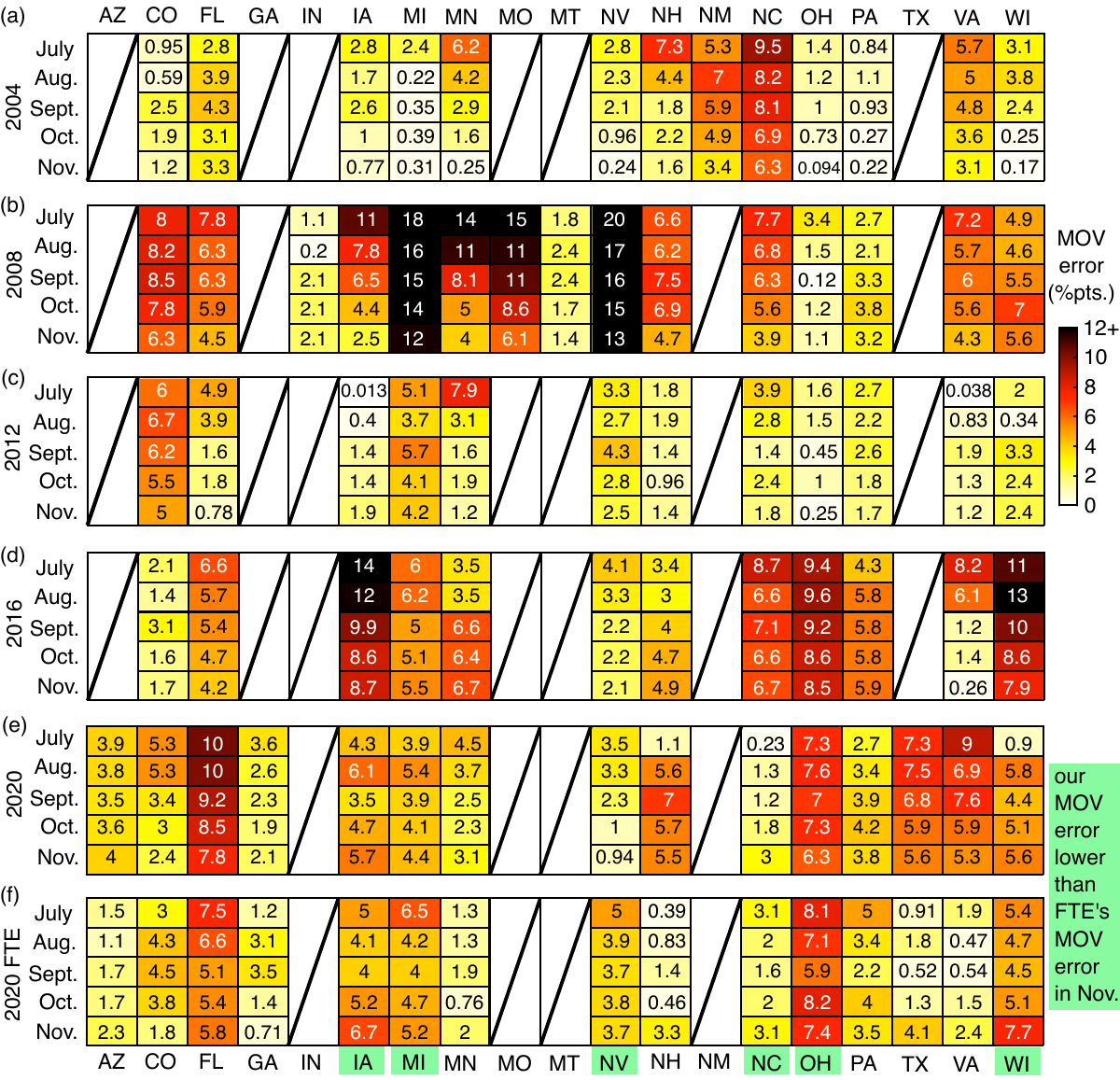}
\vspace{-\baselineskip}
\caption{Error in margin of victory in time for presidential elections in swing states using our baseline approach \cite{Volkening2020}. (a)--(e) Our MOV errors improve as we approach Election Day. (f) For reference, we show FiveThirtyEight's MOV errors \cite{FiveThirtyEight2020,538mainOld} that we compute for the $2020$ election. In (a)--(f), crossed-out boxes indicate states that are part of the Red or Blue superstates in a given year. See Figures SM2 and SM4 in the supplementary material for corresponding senatorial and gubernatorial results.}\label{fig:forecastMarginsP}
\end{figure}

Our first goal is to provide a more comprehensive picture of the behavior of the baseline approach \cite{Volkening2020} across months in a given election cycle and across elections. We show the time-dynamic ratings from our baseline forecasts of the last two decades of U.S.\ presidential elections in Fig.~\ref{fig:forecastChancesP}. As we discuss in Sect.~\ref{sec:measurement}, we forecast some regions as Red or Blue superstates, but we present ratings for all fifty states\footnote{We do not distinguish between congressional districts in Maine and Nebraska; we assign all of the electoral votes in each state to one party. FiveThirtyEight \cite{538mainNew,538mainOld}, in contrast, is able to capture the possibility of a split vote in these states. We do not take this into account when measuring their MOV error or Brier score.} and the District of Columbia individually in Fig.~\ref{fig:forecastChancesP}. Focusing on swing states, we summarize our MOV errors  for monthly presidential forecasts from July to November in Fig.~\ref{fig:forecastMarginsP}. (See Figures SM1--SM4 for the corresponding state ratings and MOV errors for senatorial and gubernatorial elections.) 

For presidential elections, our average swing-state MOV error is roughly $4.3$ \%pts.\ in October~$2020$ (about $30$ days before Election Day). Our MOV error for October $2016$ is much higher---about $5.4$ \%pts. It is also high (about $6.4$ \%pts.) 
in our forecast a month out from Election Day $2008$ due to Michigan and Nevada. Notably, our presidential MOV errors are quite low for our October $2004$ and $2012$ forecasts, both averaging $2.2$ \%pts.\ in swing states. As a whole, for presidential elections in the last two decades, our baseline approach \cite{Volkening2020} is able to capture opinion dynamics in swing states better in years with incumbents ($2004$, $2012$, and $2020$) than in years without incumbents running ($2008$ and $2016$). This echoes research suggesting that incumbency is an important feature to account for in forecasts \cite{Abramowitz1988,Campbell2014}.

\begin{figure}[t!]
\includegraphics[width=\textwidth]{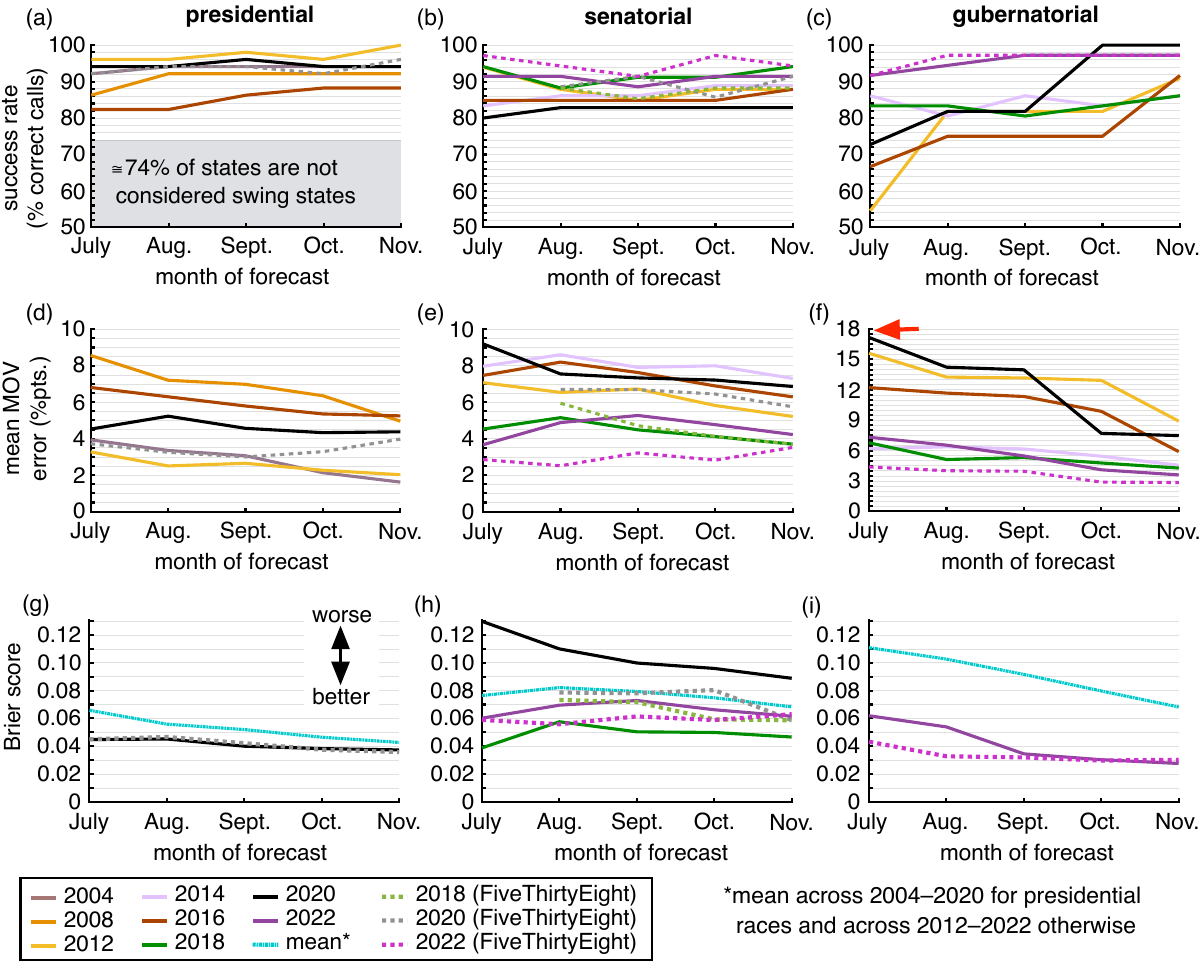}
\vspace{-\baselineskip}
\caption{Summary statistics on the performance of our baseline approach \cite{Volkening2020} for presidential elections from $2004$--$2020$ and senatorial and gubernatorial elections from $2012$--$2022$. Except for panel (f), we use the same $y$-axes across the plots in each row, to draw attention to differences based on election type. For context, we provide the summary statistics that we compute for FiveThirtyEight (based on their classic model, in cases where they provide multiple model versions) \cite{FiveThirtyEight2018,FiveThirtyEight2020,FiveThirtyEight2022} for recent elections. (a)--(c) Arguably the simplest measurement of forecast accuracy, our success rate at correctly calling election outcomes at the state level tends to improve as Election Day nears. (d)--(f) We show our average MOV errors in swing states in time for presidential, senatorial, and gubernatorial elections, respectively. (g)--(i) The Brier score is a more meaningful measurement of the accuracy of probabilistic forecasts than is success rate, and lower Brier scores indicate better forecasts \cite{Brier}. In (a)--(c) and (g)--(i), we treat each state or district that we forecast as part of a superstate as an individual; see Sect.~\ref{sec:measurement}. FiveThirtyEight \cite{FiveThirtyEight2020} did not forecast the $2020$ gubernatorial elections and, in $2018$, their data are available for forecasts from $11$ October onward, which corresponds to only one of the dates that we plot, so we do not include this point.}\label{fig:baselineMeans}
\end{figure}

In comparison to presidential elections, less research has been done on dynamically forecasting senatorial and gubernatorial elections \cite{Chen2023}. After the 2018 cycle, FiveThirtyEight \cite{538mainOld,538mainNew} also moved away from forecasting gubernatorial races. With this in mind, we provide summary statistics of the accuracy of our baseline approach \cite{Volkening2020} across presidential, senatorial, and gubernatorial elections in Fig.~\ref{fig:baselineMeans}. Whether considering swing-state MOV error or Brier score, our methods perform best for presidential elections, and we struggle the most to forecast gubernatorial races. Notably, Shirani-Mehr et al. \cite{Shirani-Mehr2018} conducted a data analysis of state-level polls for races from 1998--2014, and they found that senatorial and gubernatorial polls perform much worse than those for presidential races. This may be due to challenges in predicting turnout for senatorial and gubernatorial races \cite{Shirani-Mehr2018}. On top of this, there are fewer polls available for these races, particularly early in the election cycle \cite{Chen2023}.

Across presidential, senatorial, and gubernatorial elections, Fig.~\ref{fig:baselineMeans} also highlights that our forecasts, whether measured by success rate, mean MOV error in swing states, or Brier score, tend to improve in time. This is in agreement with observations \cite{Wlezien2004,Wlezien,Traugott2014,Campbell} that trial-heat polls improve as Election Day nears. Considering presidential elections from 1952--2000, for example, Wlezien and Erikson \cite{Wlezien2004} noted a major uptick in the performance of their regression-based forecasts in the last sixty days before Election Day. The authors \cite{Wlezien} also found that polls from 1944--2000 were much noisier and more volatile before fall, particularly during convention time in the summer.

As a point of reference before we turn to our extended approach, it is useful to discuss our results in relation to FiveThirtyEight \cite{538mainOld}, as an example of a forecaster that makes time- and pollster-specific adjustments to polls. In $2020$, FiveThirtyEight \cite{538mainOld} was about $1.2$~\%pts.\ better at capturing the true margins in swing states during the presidential race; see Fig.~\ref{fig:forecastMarginsP}(e)--(f). On the other hand, our presidential and gubernatorial forecasts in $2020$ perform comparably to FiveThirtyEight's according to success rate and Brier score. Regardless of the summary statistic in Fig.~\ref{fig:baselineMeans}, our senatorial forecasts in $2020$ and $2022$, though not in $2018$, are less accurate. Because we treat all polls equally in our baseline approach, one factor in these differences may be the adjustments that FiveThirtyEight \cite{538mainNew,538mainOld} makes to the polling data.

\subsection{Estimating historical pollster tendencies}\label{sec:extended1}

\begin{figure}[t!]
\includegraphics[width=\textwidth]{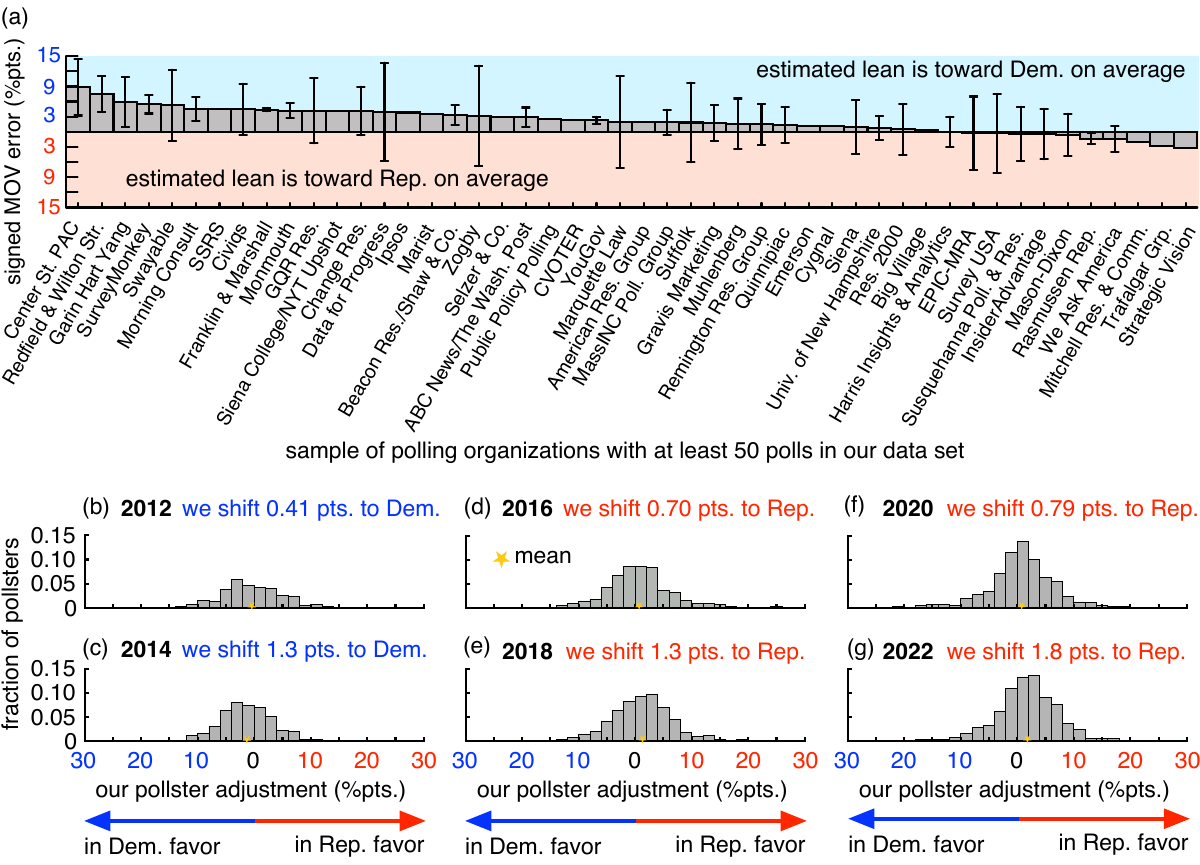}
 \vspace{-\baselineskip}
 \caption{Overview of our historical pollster-lean calculations. (a) We highlight the average historical tendency (as the mean signed MOV error in past elections according to Eqn.~\eqref{delta}) that we estimate for a sample of active polling organizations. These data are based on all of the state-level polls in our data set for presidential, senatorial, and gubernatorial elections from $2004$--$2022$. (b)--(g) As we show in Figures~\ref{fig:intro}(d) and \ref{fig:pipeline}, our extended approach involves adjusting each reported poll margin by the historical lean that we estimate for its pollster, prior to fitting parameters. We define ``historical" from the perspective of the election that we are forecasting: for example, if the goal is to forecast the $2016$ elections, we estimate the historical tendency of each pollster based only on their polls from $2004$--$2014$. We show the distribution of the pollster adjustments  that we use in forecasting elections in (b) $2012$, (c) $2014$, (d) $2016$, (e) $2018$, (f) $2020$, and (g) $2022$. For instance, based on polls for elections strictly prior to $2016$, we shift poll margins about $0.70$ \%pts.\ more Republican on average when we forecast the $2016$ elections using our extended approach. Our analysis is based on polls aggregated by FiveThirtyEight \cite{FiveThirtyEightLatestPolls,538mainOld,538mainNew,FiveThirtyEightGitHub}, RealClearPolitics \cite{RealClearPolitics2004,RealClearPolitics2008,RealClearPolling}, and HuffPost Pollster \cite{HuffPostAPI,HuffPostPollster} (collected and formatted for $2012$ and $2016$ in \cite{Gitlab_elections,Volkening2020}), and makes use of FiveThirtyEight's list of pollsters \cite{FiveThirtyEightRatings,538mainNew,538mainOld}. Identifying the pollster responsible for each poll by name across decades is difficult, and we stress that the results in this figure are based on our estimation and likely imperfect.}\label{fig:pollsterBias}
 \end{figure}

Drawing together polling data and model forecasts for elections in the past two decades provides a good opportunity to begin to investigate choices in forecasting. Using our methods in Sect.~\ref{sec:polling}, our first step toward this goal is to estimate the historical tendencies of pollsters. Specifically, we compute the mean signed error in the past polls that we attribute to each pollster according to Eqn.~\eqref{delta}. This error is year-dependent, as we consider the polls for elections from 2004 to $y-1$ when estimating historical pollster tendencies from the perspective of forecasts for elections in year $y$.

We summarize our estimated pollster adjustments $\{\Delta_y^k\}$ for different years $y$ in Fig.~\ref{fig:pollsterBias}. First, in Fig.~\ref{fig:pollsterBias}(a), we show $\Delta_\text{2023}^k$ for a selection of active pollsters that we identify, drawing on state- or district-level polls for elections in $2004$, $2008$, $2012$, $2014$, $2016$, $2018$, $2020$, and $2022$. Next, in Fig.~\ref{fig:pollsterBias}(b)--(g), we present distributions of our pollster adjustments $\{\Delta_y^k\}$ for elections in different years. For example, based on the historical performance of polling organizations from $2004$ to $2020$, we estimate that the polls will lean overly Democratic in $2022$, and we thus shift the poll margins toward the Republican candidate by $1.8$ \%pts.\ on average in our extended approach; see Fig.~\ref{fig:pollsterBias}(f). Notably, our mean pollster adjustment changes from a shift in favor of Democratic candidates in $2012$ and $2014$ to a shift towards Republican candidates from $2016$--$2022$. Because the amount of historical data on which to evaluate pollsters grows with each election, our pollster adjustments are based on much less data for elections in $2012$ and $2014$, as compared with races in $2018$--$2022$; see Fig.~\ref{fig:data}(a).

As we discuss in Sect.~\ref{sec:polling}, estimating the historical tendencies of pollsters is challenging due to ambiguities and inconsistencies in organization names. Thus, we fully expect that our library of pollster names is not perfect due to these challenges and the limitations of our study. However, it is notable that we identify
$481$ polling organizations active in state-level polling for elections from 2004--2022, and FiveThirtyEight's pollster ratings for $2021$ (but updated last year, in $2023$) list $492$ pollsters \cite{FiveThirtyEightGitHub}. While we caution that it is not a direct comparison, as FiveThirtyEight's data sets and our data sets are not the same, the similarity in these pollster counts is encouraging and suggests our process is producing results in the right ballpark. We provide our library of pollster-name associations on GitLab \cite{Gitlab_elections_v2}, and we stress that it is our imperfect estimation. 

While mainstream sources like FiveThirtyEight \cite{538mainOld,538mainNew} and \textit{The New York Times} \cite{NYT} account for house effects or tendencies in their forecasts, the related academic research has mainly focused on house effects in a small subset of pollsters or has estimated bias at the poll level (to our knowledge). For example, Shirani-Mehr et al.\ \cite{Shirani-Mehr2018} conducted a study of polls from 1998--2014 to estimate and distinguish between error due to house effects and error due to sample variance. Their results \cite{Shirani-Mehr2018} are presented as distributions across all polls (similar to Fig.~\ref{fig:data}(b)), as opposed to distributions across all pollsters (like Fig.~\ref{fig:pollsterBias}(b)--(g)), and they account for house effects through a variance term that depends on each poll. Forsberg and Payton \cite{Forsberg2015} provide estimates of bias in the polls from the last ten days before the 2004, 2008, and 2012 presidential elections, considering about twenty pollsters. Our work with pollster history involves many simplifications in this first study, but we see our large-scale, comprehensive approach to the polls as a contribution. We hope that our pollster-name associations on GitLab \cite{Gitlab_elections_v2} provides a useful starting point from which others can continue to build.

\subsection{A study of how accounting for historical pollster tendencies affects forecasts}\label{sec:extended2}

\begin{figure}[t!]
\includegraphics[width=\textwidth]{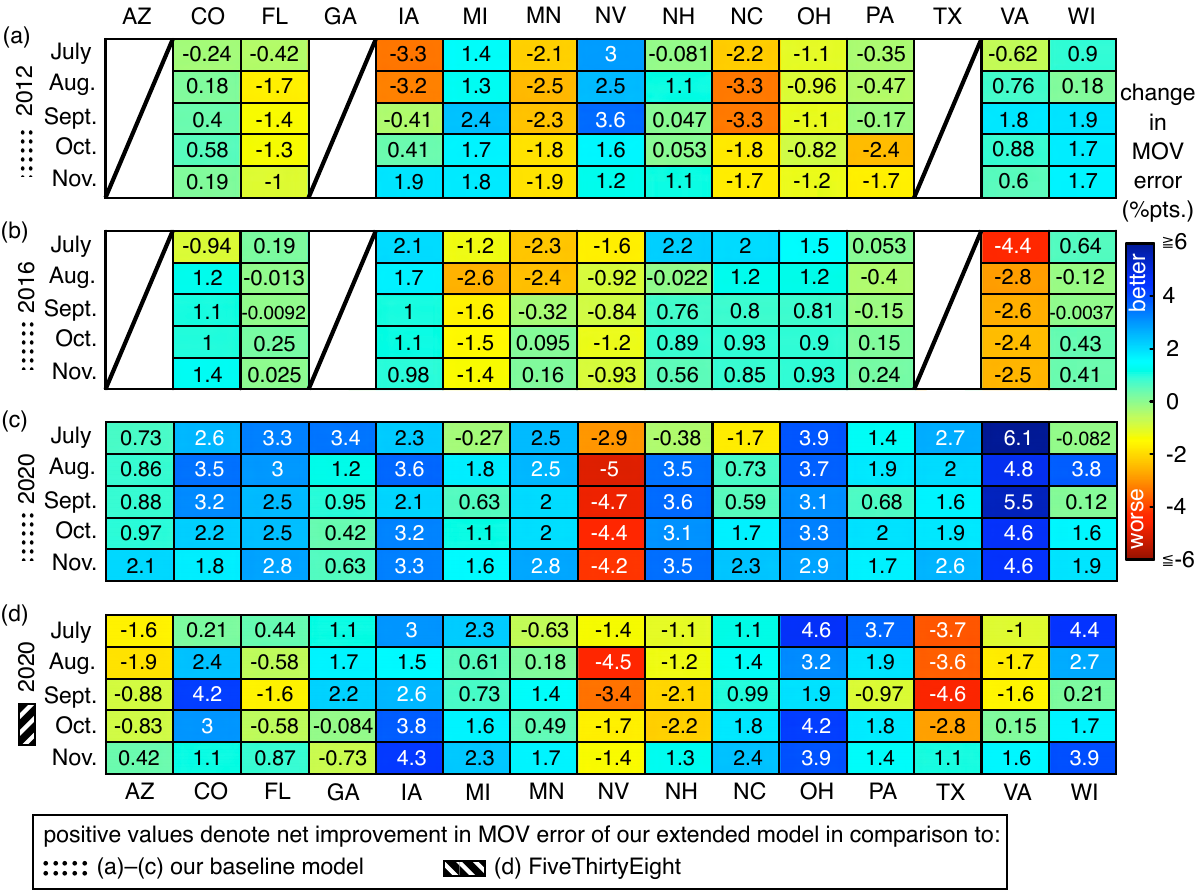}
\vspace{-\baselineskip}
\caption{Difference in MOV error in swing states for our baseline and extended approaches applied to presidential elections. We show the MOV errors from our baseline approach \cite{Volkening2020} minus the MOV errors from our extended approach for (a) $2012$, (b) $2016$, and (c) $2020$, so positive numbers indicate that accounting for pollster partisan lean improves our forecasts. (d) For reference, we also show the difference between the MOV errors that we compute for FiveThirtEight \cite{FiveThirtyEight2020} and the MOV errors from our extended approach in $2020$. Notably, in panels (c)--(d) our extended approach results in substantial improvement in $2020$ for many swing states, but not all---for example, accounting for pollster partisan lean worsens our forecasts for Nevada. Also see Figures SM5--SM6.\label{fig:MOVp}}
\end{figure}

With our estimates of pollster historical tendencies in hand (Fig.~\ref{fig:pollsterBias}), we now investigate how one way of making pollster-specific adjustments to the polls affects our forecast accuracy for elections from $2012$--$2022$. Focusing on MOV error in swing states, we show how our extended approach performs relative to our baseline approach \cite{Volkening2020} for presidential elections in Fig.~\ref{fig:MOVp}. There are presidential forecasts available from FiveThirtyEight \cite{538mainOld} in time for the $2020$ elections, and we compare our extended approach to these forecasts in Fig.~\ref{fig:MOVp}(d) to provide more context. (Figures SM$5$--SM$6$ present our corresponding senatorial and gubernatorial results.) Across Fig.~\ref{fig:MOVp} and Figures~SM$5$--SM$6$, positive numbers and blue-ish tones indicate that accounting for the historical tendencies of pollsters improves forecast accuracy. Notably, our MOV error for the $2020$ presidential election drops very strongly when we apply our extended approach, improving our performance in all swing states except Nevada.

\begin{figure}[t!]
\includegraphics[width=\textwidth]{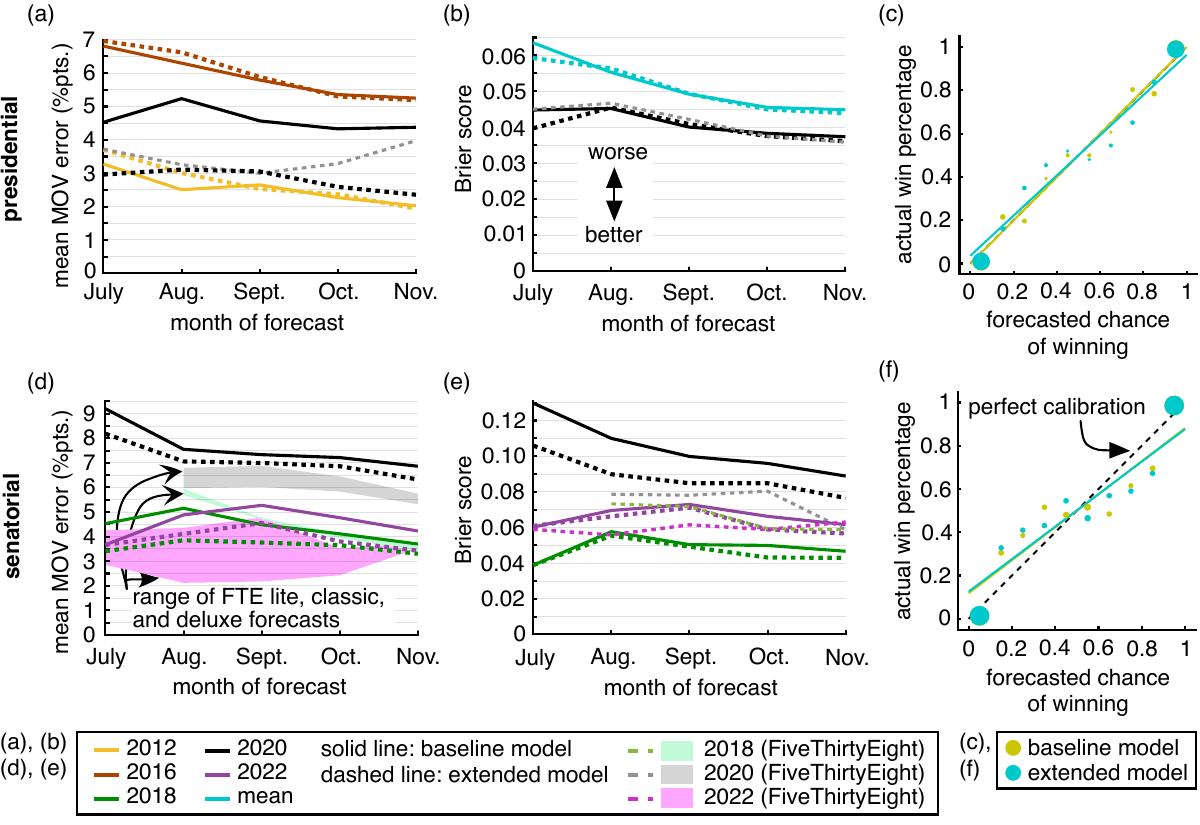}
\vspace{-\baselineskip}
\caption{Summary statistics on the performance of our baseline \cite{Volkening2020} and extended approaches for (a)--(c) presidential and (d)--(f) senatorial elections from $2012$ to $2022$. As a reference, we include the summary statistics that we compute for FiveThirtyEight \cite{FiveThirtyEight2018,FiveThirtyEight2020,FiveThirtyEight2022} for recent elections. (a) Our two approaches produce similar mean MOV errors in swing states for the $2012$ and $2016$ presidential elections; however, we see a large improvement from our extended approach for the $2020$ presidential elections. We highlight that our extended approach leads to lower mean MOV errors in swing states than does FiveThirtyEight \cite{FiveThirtyEight2020} from September $2020$ onward. Importantly, our $2012$ adjustments for pollster history are based on $2004$ and $2008$ data, and our $2016$ adjustments are based on $2004$--$2014$ data; our extended approach performs particularly poorly for $2014$; see Fig.\ SM7. (b) Lower Brier scores correspond to better probabilistic forecasts. We show our mean Brier scores (calculated using all of the states that we forecast) for presidential elections from $2012$--$2020$, as well as our $2020$ Brier score; we also include FiveThirtyEight's $2020$ presidential Brier score that we compute. (c) In terms of calibration for our presidential forecasts from July through November $2012$--$2020$, we plot $(x,y)$ points, where $y$ is the fraction of times we give a candidate a $p \in [x-\Delta x/2,x +\Delta x/2)$ percent chance of winning and that candidate wins. We also show the best-fit lines to these data. (d)--(f) We include the corresponding summary statistics for our baseline and extended approaches for senatorial elections. In (d), the shaded regions denote the range of FiveThirtyEight's forecasts, according to the three different versions of their methods for senatorial elections from $2018$--$2022$ \cite{FiveThirtyEight2018,FiveThirtyEight2020,FiveThirtyEight2022}. See Fig.~SM7 for corresponding gubernatorial summary statistics.\label{fig:summary2}}
\end{figure}

Figure~\ref{fig:summary2} presents summary statistics on our presidential and senatorial forecasts, comparing our baseline \cite{Volkening2020} and extended approaches. (See Fig.~SM$7$ for the summary statistics for gubernatorial elections.) The strong improvement in our $2020$ presidential forecasts stands out again in these summary statistics. For example, incorporating pollster history into our methods improves our mean  MOV error in swing states by about $1.8$ \%pts., averaging across our monthly forecasts. Moreover, when we compare $2020$ presidential forecasts from our extended approach and FiveThirtyEight \cite{538mainOld,538mainNew,FiveThirtyEight2020}, our mean swing-state MOV error is $0.64$ \%pts.\ lower on average than FiveThirtyEight's errors for forecasts on the same five dates. 

Starting with the $2018$ elections, FiveThirtyEight \cite{538mainOld,FiveThirtyEight2018,538mainNew} gave viewers the choice to view the ``lite", ``classic", or ``deluxe" versions of their forecasts. The lite forecast is polls-driven, the classic forecast combines polls with fundamental data, and the deluxe forecast merges FiveThirtyEight's classic forecast with other analysts' opinions \cite{538mainOld,FiveThirtyEight2018}. The FiveThirtyEight team continued this approach of lite, classic, and deluxe forecasts for their senatorial and gubernatorial forecasts in $2020$ and $2022$, but they limited their presidential forecasts to a single version in $2020$ \cite{538mainOld,FiveThirtyEight2020}. We provide the range that the mean MOV errors in swing states occupy in time based on these three approaches as a shaded region for each senatorial election in Fig.~\ref{fig:summary2}(d). Interestingly, there has been a gradual increase in the size of these regions, indicating that the difference between FiveThirtyEight's \cite{538mainOld} lite and deluxe forecasts of the senatorial elections has expanded from $2018$ to $2022$. In the case of the $2018$ senatorial elections, our baseline approach and all of FiveThirtyEight's models \cite{538mainOld} produce similar mean errors in swing states; our extended approach, in turn, leads to slightly more accurate forecasts across time. While both our baseline and extended approaches produce higher $2020$ and $2022$ MOV errors than do nearly all of FiveThirtyEight's models \cite{538mainOld} in Fig.~\ref{fig:summary2}(d), accounting for pollster historical performance reduces this difference in performance some.

Taking our results in Figures~\ref{fig:MOVp}--\ref{fig:summary2} and Figures~SM$5$--SM$7$ as a whole, we conclude that adjusting for the historical tendencies of pollsters generally improves our presidential and senatorial forecasts in $2018$, $2022$, and $2022$. Both our baseline and extended approaches perform poorly for gubernatorial forecasts, and we suggest that accounting for pollster history according to our methodology does not improve forecasts in the case of races for governor positions. Viewing our results in Fig.~\ref{fig:MOVp} alongside Fig.~\ref{fig:forecastMarginsP}(e)--(d) also suggests a few cautionary comments. While we see improvements of $2$--$3$ \%pts.\ in our $2020$ presidential forecasts in many swing states under our extended approach, 
accounting for historical pollster tendencies causes our MOV error in Nevada to increase by over $4$ \%pts.\ in our late $2020$ forecasts. Moreover, as we discuss in Fig.~SM$7$, our extended approach performs almost $2$ \%ps.\ worse on average in swing states for the $2014$ senatorial elections, despite out-performing our baseline approach in all other senatorial races from $2012$--$2022$. This highlights the major challenge with accounting for the historical tendencies of pollsters: the process of adjusting for past pollster performance inherently relies on historical data. Past elections and pollster performance need not be predictive of future elections, as pollsters likely change their methods in time \cite{Selb2016,Selb2023}.

\section{Conclusions and discussion}\label{sec:conclusions}

\begin{figure}[t!]
\centering
\includegraphics[width=\textwidth]{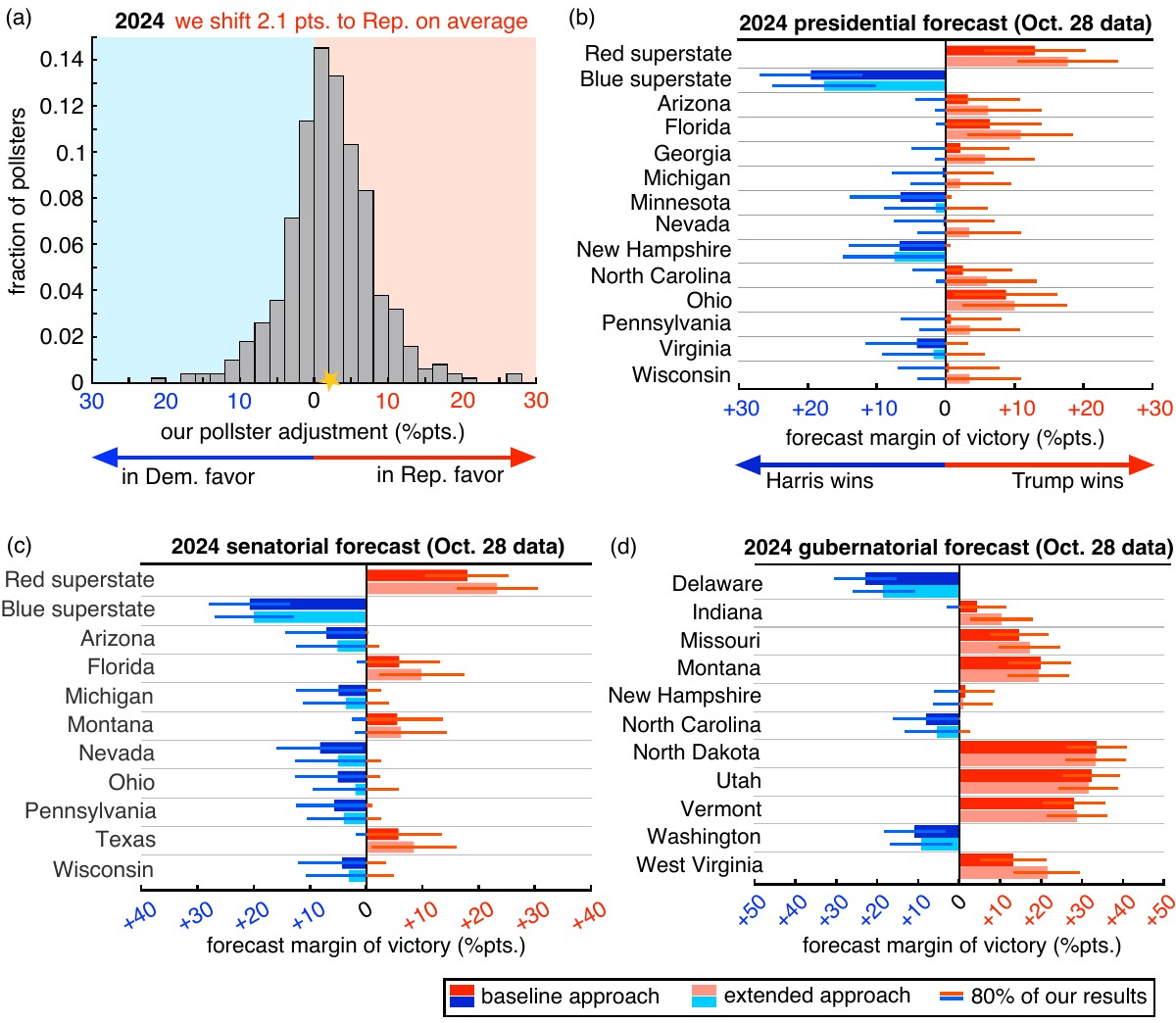}
\vspace{-\baselineskip}
\caption{Discussion of our $2024$ forecasts. See our website \cite{2024forecasts} for forecasts in time. (a) The pollster adjustments that we use in our extended approach to forecasting the $2024$ elections lean more Republican than in all of the past years that we consider; see Fig.~\ref{fig:pollsterBias}. As we discuss in Sect.~\ref{sec:extended1}, these adjustments are based on our estimate of the average historical signed MOV error in state-level polls across pollsters in our data set. Here we use polls from $2004$--$2022$. (b) Based on polling data that we downloaded from FiveThirtyEight \cite{FiveThirtyEightLatestPolls,538mainNew} on $28$ October, we present our forecasts of the presidential elections; also see our website \cite{2024forecasts}. Our extended approach that accounts for pollster history in a simplified way in this initial study shifts our presidential forecasts more in favor of the Republican candidate. The bars indicating where the middle $80$\% of our simulated election results fall highlight the closeness of the $2024$ election; for most swing states, our approaches consider both Democratic and Republican outcomes possible. We also include our forecasts of the (c) senatorial and (d) gubernatorial elections based on data that we downloaded on $28$ October from FiveThirtyEight\cite{FiveThirtyEightLatestPolls,538mainNew}. We recommend the recent studies \cite{Yang2024,Gelman2020} for some discussion of considerations when viewing and evaluating probabilistic forecasts.\label{fig:2024}}
\end{figure}

We completed a comprehensive study of the compartmental Republican--Undecided--Democratic model \cite{Volkening2020}, presenting monthly forecasts of the presidential, senatorial, and gubernatorial elections from $2004$--$2022$. This allowed us to illustrate how the performance of the cRUD model \cite{Volkening2020} depends on the time before Election Day and to put our forecasts into context by viewing results across $20$ years. Addressing the challenges presented by differences in pollster-naming conventions across data sources and time, we estimated the historical tendencies of each pollster in our data set. With our library of about $480$ estimated polling organizations (available on GitLab \cite{Gitlab_elections_v2}), we developed an extended approach to forecasting that involves shifting each poll's margin based on the average historical lean of its pollster before fitting model parameters. We discussed how this approach to incorporating house effects into forecasts impacts accuracy, and we hope our initial analysis encourages future work.

Considering monthly forecasts across elections, we found that our forecast accuracy reflects observations that poll errors decrease as Election Day nears \cite{Wlezien2004,Wlezien,Traugott2014,Campbell} and tend to be larger for senatorial and gubernatorial elections \cite{Chen2023,Shirani-Mehr2018}. We also observed that our presidential forecasts were more accurate in years with an incumbent running, echoing studies \cite{Wlezien2004,Wlezien,Traugott2014,Campbell,Campbell2014} that suggest incumbency is important. Looking ahead to the $2024$ presidential election, it is thus notable that this race features essentially two incumbents, Harris (a member of the incumbent party) and Trump (a prior president). For presidential elections in $2012$ and $2016$, we found little difference in the performance of our baseline and extended approaches. This may be due to the much lower number of polls on which we are evaluating pollster performance for these years; see Fig.~\ref{fig:data}(a). In the case of the $2020$ presidential elections, on the other hand, we observed a large improvement when we accounted for pollster history. We also saw strong improvements in our $2018$, $2020$, and $2022$ senatorial forecasts when we accounted for house effects. However, there are exceptions to these observations: accounting for pollster history hurt our performance in some states. Moreover, the $2014$ elections reacted very differently to our extended approach than did the other races from $2012$--$2022$.

With the $2024$ presidential, senatorial, and gubernatorial elections in mind, a natural next question is what our baseline \cite{Volkening2020} and extended approaches have to say about these races. The dynamics with Biden leaving the presidential race and Harris entering it are unusual, and we faced choices on how to account for this. In the end, we chose to restrict to polls of Harris vs.\ Trump to estimate model parameters. We provide these forecasts, as well as the distribution of pollster adjustments that we estimate based on the polls in our data set from 2004--2022, in Fig.~\ref{fig:2024}. 
As we illustrate in Fig.~\ref{fig:2024}(a), the pollster shifts that we estimate for $2024$ are our largest yet, with a mean shift of $2.1$ \%pts.\ toward Republican candidates. We stress, however, that these estimated shifts are fully based on the historical lean that we identify for each pollster, and---as we discuss throughout this study---opinion dynamics and forecast accuracy change in time. With this word of caution in mind, we consider the $2024$ elections a useful means of further testing choices in working with polling data and suggesting future directions.

There are many places to improve and build on our underlying cRUD model, forecasting approach, and means of accounting for house effects. For example, our approach to estimating the historical tendency of each pollster as their mean signed MOV error is a simplification. In this process, we are assuming that a pollster's errors due to sampling cancel out across their polls, and what remains is their error due to house effects. For pollsters with many polls, this simplification may make sense, but it is less reasonable to apply it to pollsters with only a few polls. Perhaps a better approach would be to distinguish sampling variance from poll bias, as in \cite{Shirani-Mehr2018,deStefano2022}. In terms of the resulting measurements of pollster tendencies and their effects on forecast accuracy, it would also be interesting to compare how our approach (relying on history) compares to other methods for estimating house effects. These could include accounting for the methodologies of pollsters (when they are not proprietary), using polls of the current election to estimate house effects under the assumption that there is no net bias across the polls, or blending these ideas together \cite{Pickup2008}.

From a dynamical-systems and multiscale-modeling perspective, we see exciting opportunities for future work that involve nonlinear dynamics across and within election cycles. In particular, a central challenge in adjusting polling data is that it is difficult to predict what will cause the next major polling miss \cite{Gelman2020}. Will the polls miss the true margins because of late deciders, mis-idenfication of likely voters, or something else? In the case of the German elections, Selb et al.\ \cite{Selb2016} found an oscillating pattern in poll errors: pollsters appeared to collectively react to each election by over-adjusting their methods in the next election to correct for their recent mistakes. Dynamical systems-based methods and models are well-suited to capturing such patterns. In the future, we suggest it would be very interesting to combine a within-cycle model of voter dynamics with an across-cycle model of how the polls should be adjusted and interpreted. As another direction, at the shorter timescale of an election campaign, it is notable that a person's chance of responding to a poll may depend on whether their preferred candidate is ahead \cite{Gelman2021}. On top of this, studies have suggested that viewing forecasts or surveys can influence a person's chance of turning out to vote \cite{Westwood,turnout}. Again, these are places where nonlinear dynamics and feedback may be at work during elections.

One of reasons that we chose to take a historical perspective on pollster tendencies is because we see associating the various strings that refer to a given pollster's name across decades and data sources as opening up many directions for future work. One of these is related to FiveThirtyEight's ``pollster ratings" for weighting polls \cite{538mainNew,538mainOld}. In comparison to adjusting the polls to account for partisan lean, designing a transparent method for ranking pollsters would allow one to more heavily weight polls from organizations that are judged as better. There are many other pollster- and time-specific adjustments that can be made when working with polls \cite{538mainOld,538mainNew}, and it would be interesting to determine how each of theses adjustments affects forecasts. In addition to testing out these adjustments in the cRUD model, it will be important to see if the results are consistent across different modeling approaches (such as the voter model \cite{Fernandez2014}) and parameter-estimation approaches.

Another exciting direction would be to consider the dynamics of split-ticket voting across decades \cite{Cook2014}. Perhaps information from presidential elections could be used to inform senatorial and gubernatorial forecasts, increasing accuracy. Wang also suggested that presidential and senatorial elections move together \cite{Wang}. With the 2024 election in mind, one could ask about the role of third party votes and so-called ``spoiler candidates" by adding more differential equations to the cRUD model. Related to this, it would be interesting to extend our forecasting approach to other countries (perhaps with more candidates running) in which polling data are publicly available. More broadly, we encourage folks to build on the initial study \cite{Volkening2020} and our second step here, as there are a wealth of election-related questions into which we expect an applied-dynamical systems perspective could provide insight.

\section*{Code and data availability} \label{sec:data}

The \textsc{Matlab} and \textsc{R} programs that we developed in support of this study, as well as reproducibility instructions, are publicly available on GitLab \cite{Gitlab_elections_v2}. We also provide our formatted data or instructions on how to download these data from the original sources. See \cite{2020forecasts,2022forecasts,2024forecasts} for the $2020$, $2022$, and $2024$ websites associated with the real-time forecasts generated using the methods in our study and the original study \cite{Volkening2020}.

\appendix

\section{Correlating noise} \label{sec:appJaccard}
Our approach to correlating noise in Eqns.~\eqref{eq:sde1}--\eqref{eq:sde2} is the same as the one in \cite{Volkening2020}, but with updated sources of demographic data. Specifically, we quantify the similarity of regions $i$ and $j$ by computing $J_{i,j} = \min\{X^i,X^j\}/\max\{X^i,X^j\}$, where $X^i$ and $X^j$ denote the fractions of some demographic of interest in regions $i$ and $j$, respectively. As in \cite{Volkening2020}, we compute three Jaccard indices using different demographic data: $[J_{i,j}^\text{E}]$, $[J_{i,j}^\text{B}]$, and $[J_{i,j}^\text{H}]$, for $1\le i,j, \le M$. We use demographic data on education levels (for $[J_{i,j}^\text{E}]$) from the U.S.\ Census Bureau \cite{EducationData2020201820162014201220082004} for elections from 2004--2020 and from the Federal Reserve Bank of St.\ Louis\cite{EducationData2022,EducationData2024} for elections in 2022 and 2024. We use state-level demographic data on the fraction of non-Hispanic Black individuals (for $[J_{i,j}^\text{B}]$) and on the fraction of Hispanic individuals (for $[J_{i,j}^\text{H}]$) from the U.S.\ Census Bureau for 2004--2020 \cite{DemographicData2020201820162014201220082004}, 2022 \cite{DemographicData2022}, and 2024 \cite{DemographicData2024}. For each simulation of Eqns.~\eqref{eq:sde1}--\eqref{eq:sde2}, we uniformly at random select one of these three Jaccard indices and then use it as the covariance for $d\textbf{W}_\text{R}(t)$ and $d\textbf{W}_\text{D}(t)$, which have multivariate normal distributions with mean zero. See \cite{Volkening2020} for details.

\section{Numerical implementation}\label{sec:appNumerical}
We follow the approach \cite{Volkening2020} to numerically implement our methods. Briefly, we format polling data and simulate Eqns.~\eqref{eq:sde1}--\eqref{eq:sde2} in \textsc{Matlab}. We use the Euler--Maruyama method \cite{Higham} to solve Eqns.~\eqref{eq:sde1}--\eqref{eq:sde2} numerically with a time step of $\Delta t = 1$ day and noise strength of $\sigma = 0.0015$ from 1 January to Election Day of the same year. Because we reduce our system to two equations per region $i$ using that $R_i(t) + U_i(T) + D_i(t) = 1$, we conserve the total population in time. Our parameter-fitting approach follows the one in \cite{Volkening2020}: we perform optimization of our objective function with constraints for non-negativity using the \textsc{OPTIM} function in \textsc{R} \cite{R,byrd1995limited} with a time step of $\delta t = 3$ days for $T$ months. As in \cite{Volkening2020}, we approximate every month as $30$ days long when fitting parameters and simulating opinion dynamics. For example, because Election Day 2022 was 8 November, we simulate Eqns.~\eqref{eq:sde1}--\eqref{eq:sde2} for $308$ days to produce our forecasts.

\section{Supplementary material}

\setcounter{figure}{0}
\renewcommand{\figurename}{Figure}
\renewcommand{\thefigure}{SM\arabic{figure}}

This supplementary material contains the following:
\begin{itemize}[noitemsep,nolistsep]
\item Sect.~\ref{sec:addresults}: additional sen.\ and gub.\ results using our baseline approach;
\item Sect.~\ref{sec:addresults2}: additional sen.\ and gub.\ results using our extended approach; and
\item Sect.~\ref{sec:superstates}: superstate definitions for each election.
\end{itemize}

\subsection{Additional senatorial and gubernatorial results}\label{sec:addresultsO}

The main manuscript focuses on presidential elections and briefly discusses senatorial and gubernatorial races. For completeness, here we provide more detail on our forecasts---first using our baseline approach and then using our extended approach---for elections of senators and governors from $2012$--$2022$.

\subsubsection{Additional sen.\ and gub.\ results using our baseline approach}\label{sec:addresults}

\begin{figure}[t!]
\includegraphics[width=\textwidth]{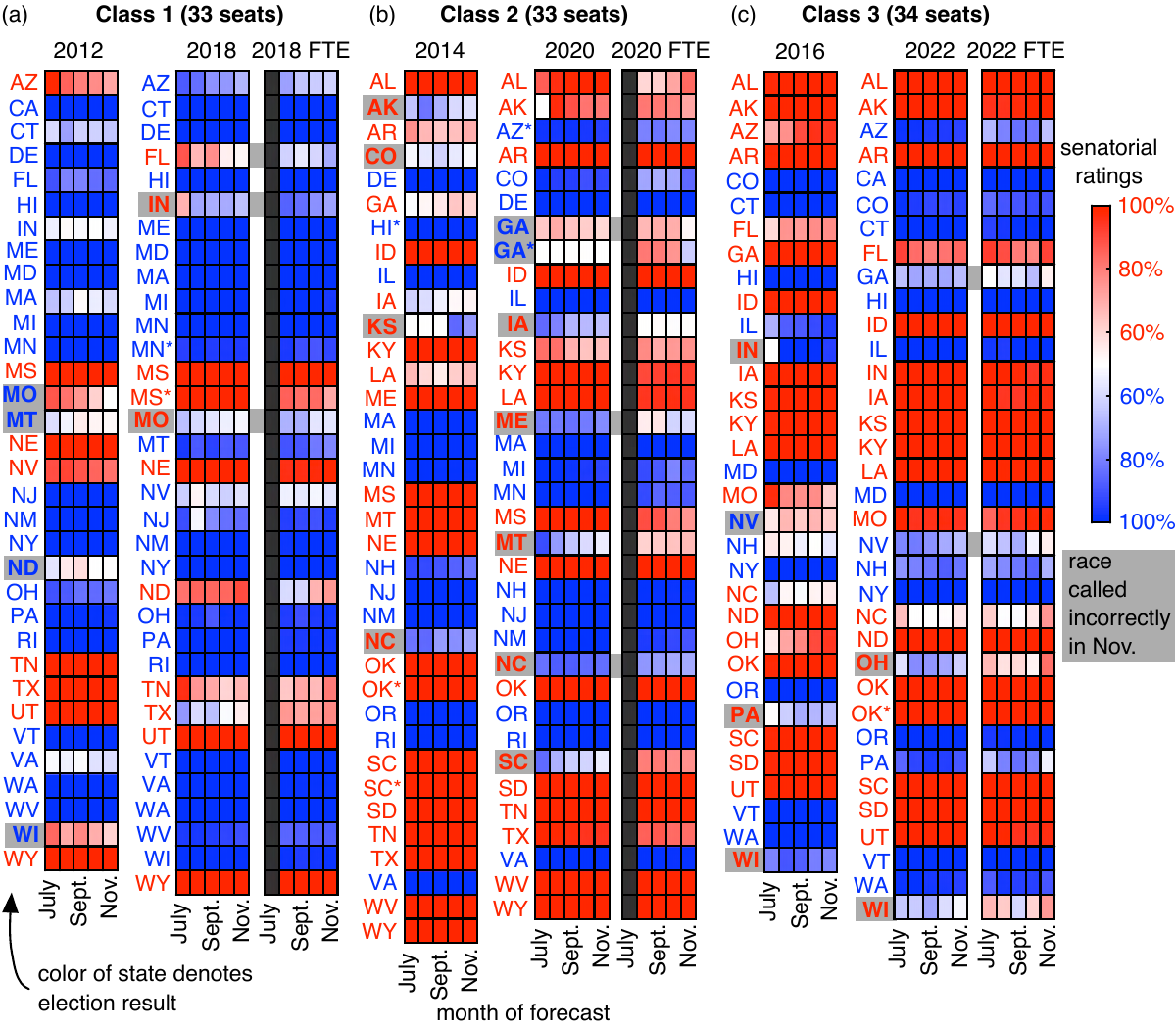}
\vspace{-\baselineskip}
\caption{Ratings in time using our baseline approach to forecast senatorial elections between $2012
$--$2022$. There are $100$ seats in the U.S.\ Senate, and they are broken up into three classes that rotate holding elections. We show the ratings that we generate from our baseline approach each month from July to November for senatorial races in (a) $2012$ and $2018$ (Class $1$), (b) $2014$ and $2018$ (Class $2$), and (c) $2016$ and $2022$ (Class $3$). In (b) and (c), we also provide some ratings from FiveThirtyEight (classic version) \cite{FiveThirtyEight2018,FiveThirtyEight2020,FiveThirtyEight2022} for reference and context. Following the same approach that we use in Fig.~$4$ in the main manuscript, we show ratings for every state individually, although some states are forecast as part of our Red or Blue superstates; see Sect.~\ref{sec:superstates} for superstate definitions. We denote election outcomes by the color of the state abbreviation, and we indicate state races that we call incorrectly in gray. \label{fig:senRatings}}
\end{figure}

\begin{figure}[t!]
\centering
\includegraphics[width=0.77\textwidth]
{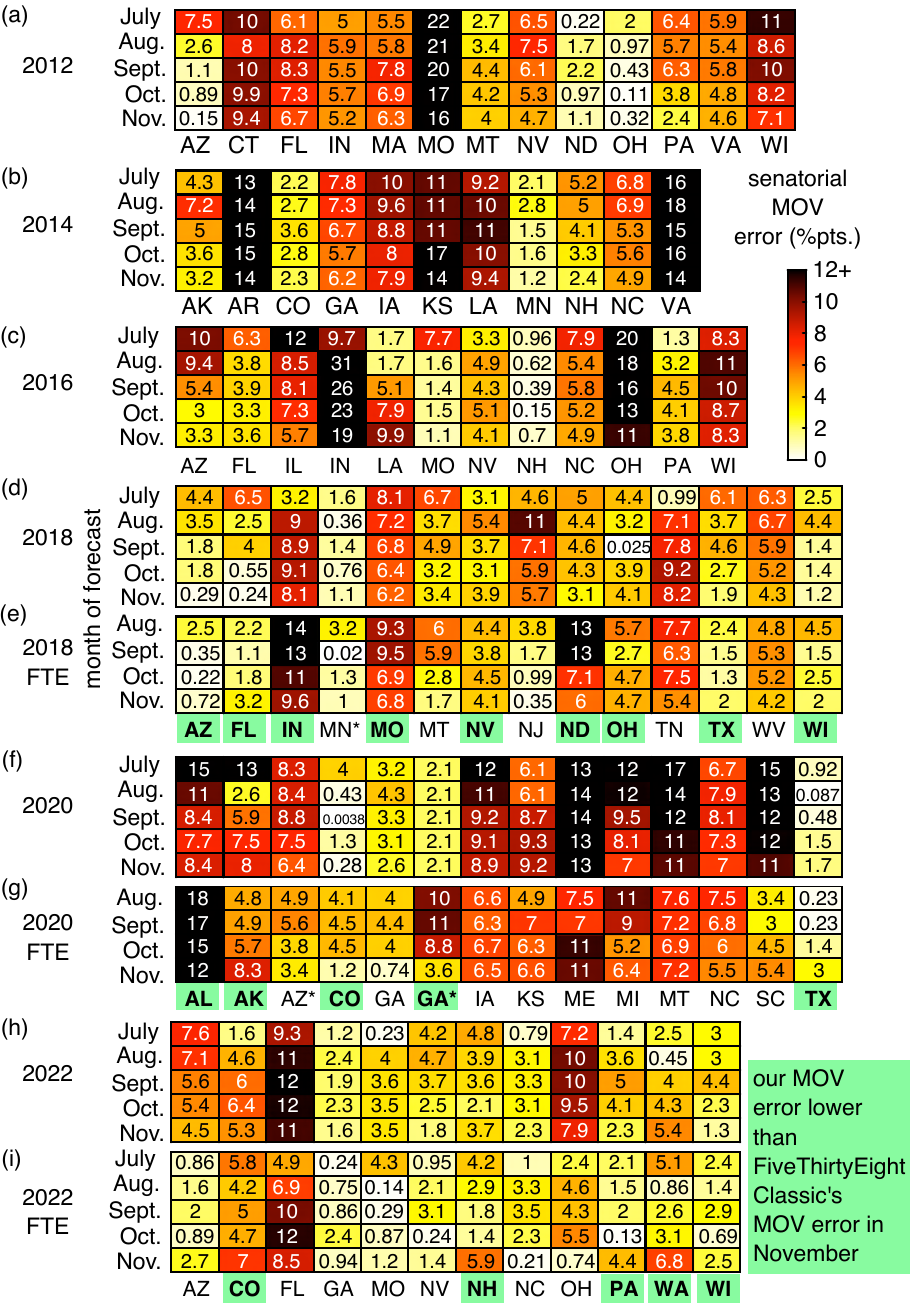}
\caption{MOV errors in time for forecasts of senatorial elections in swing states. We show the MOV errors that we produce using our baseline approach for senatorial elections in (a) 2012, (b) 2014, (c) 2016, (d) 2018, (f) 2020, and (h) 2022. Additionally, we show the MOV errors that we compute for FiveThirtyEight's senatorial forecasts (classic version) in (e) 2018, (g) 2020, and (i) 2022 for reference \cite{FiveThirtyEight2018,FiveThirtyEight2020,FiveThirtyEight2022}. We indicate state races for which the MOV error from our baseline approach is lower (i.e., better) than that from FiveThirtyEight \cite{538mainOld} in green. We round all MOV errors and display two significant figures. \label{fig:senMOV}}
\end{figure}

In Figures~\ref{fig:senRatings} and \ref{fig:senMOV}, respectively, we show the state ratings and MOV errors that we find using our baseline approach for senatorial elections. Notably, we call two senatorial election outcomes incorrectly for 2018: Indiana and Missouri, one less than FiveThirtyEight \cite{FiveThirtyEight2018}, which incorrectly leaned more Democratic for Florida. Figure~\ref{fig:senMOV}(a) further shows that our baseline approach produces lower MOV errors for $2018$ than does FiveThirtyEight's classic model \cite{FiveThirtyEight2018}. In comparison, for 2020 and $2022$, we perform worse than FiveThirtyEight by both success rate and MOV error.

\begin{figure}[t!]
\includegraphics[width=\textwidth]{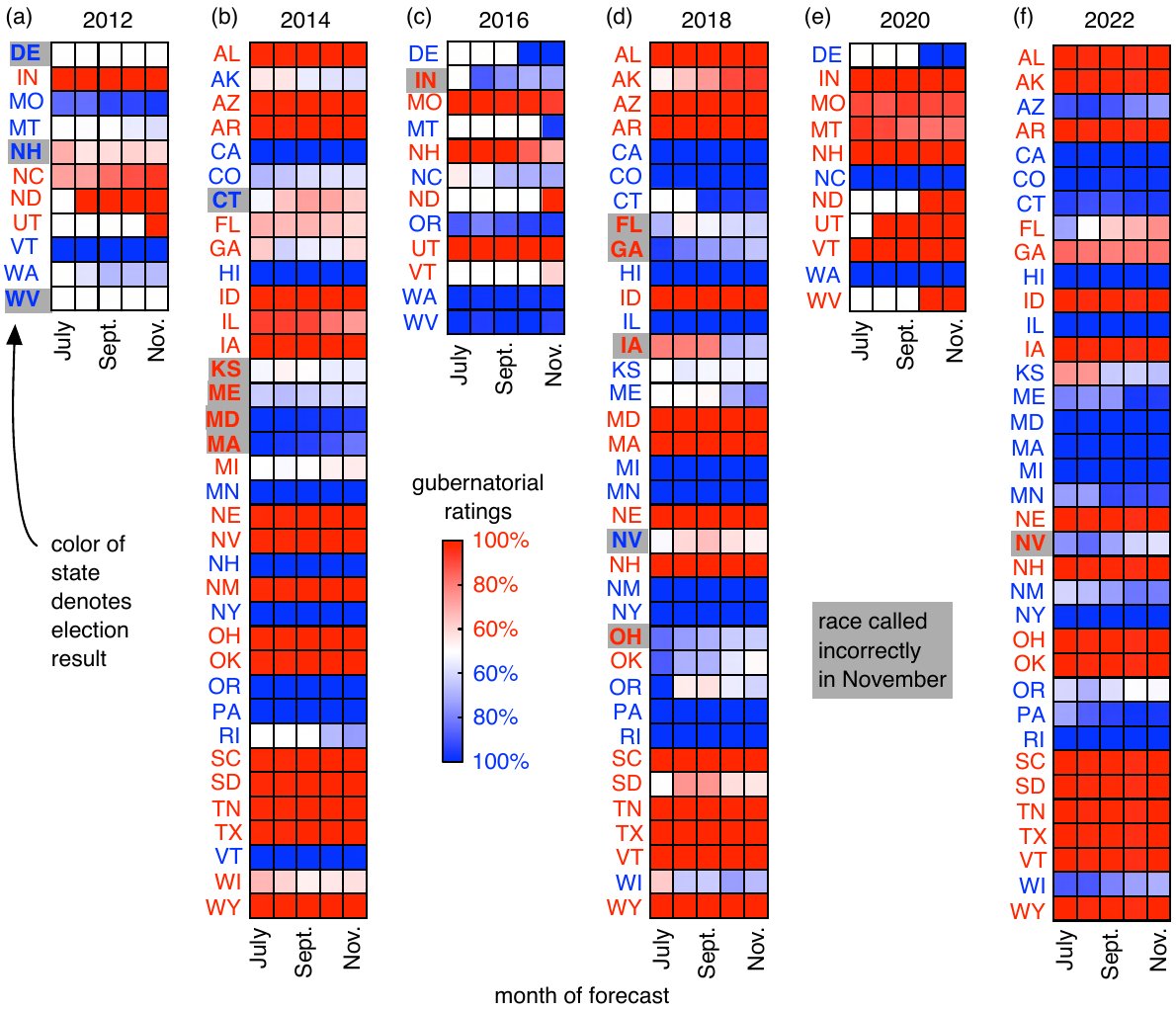}
\vspace{-\baselineskip}
\caption{Ratings in time using our baseline approach to forecast gubernatorial elections each month from July through November. We show ratings for gubernatorial elections in (a) $2012$, (b) $2014$, (c) $2016$, (d) $2018$, (e) $2020$, and (f) $2022$. We indicate election outcomes by the text color of the state abbreviations, and highlight races that we called incorrectly in gray. While we combine some states into Red or Blue superstates for $2014$, $2018$, and $2022$ (see Sect.~\ref{sec:superstates}), we show state ratings for all states individually here. Less gubernatorial polling data are available compared to senatorial and presidential polls, and, in some cases, no polls are available from our data sources for states that we forecast individually. When this occurs (i.e., see Delaware in $2012$ and early $2020$), we report the race as a toss-up.\label{fig:gubRatings}}
\end{figure}

\begin{figure}[t!]
\centering
\includegraphics[width=0.875\textwidth]{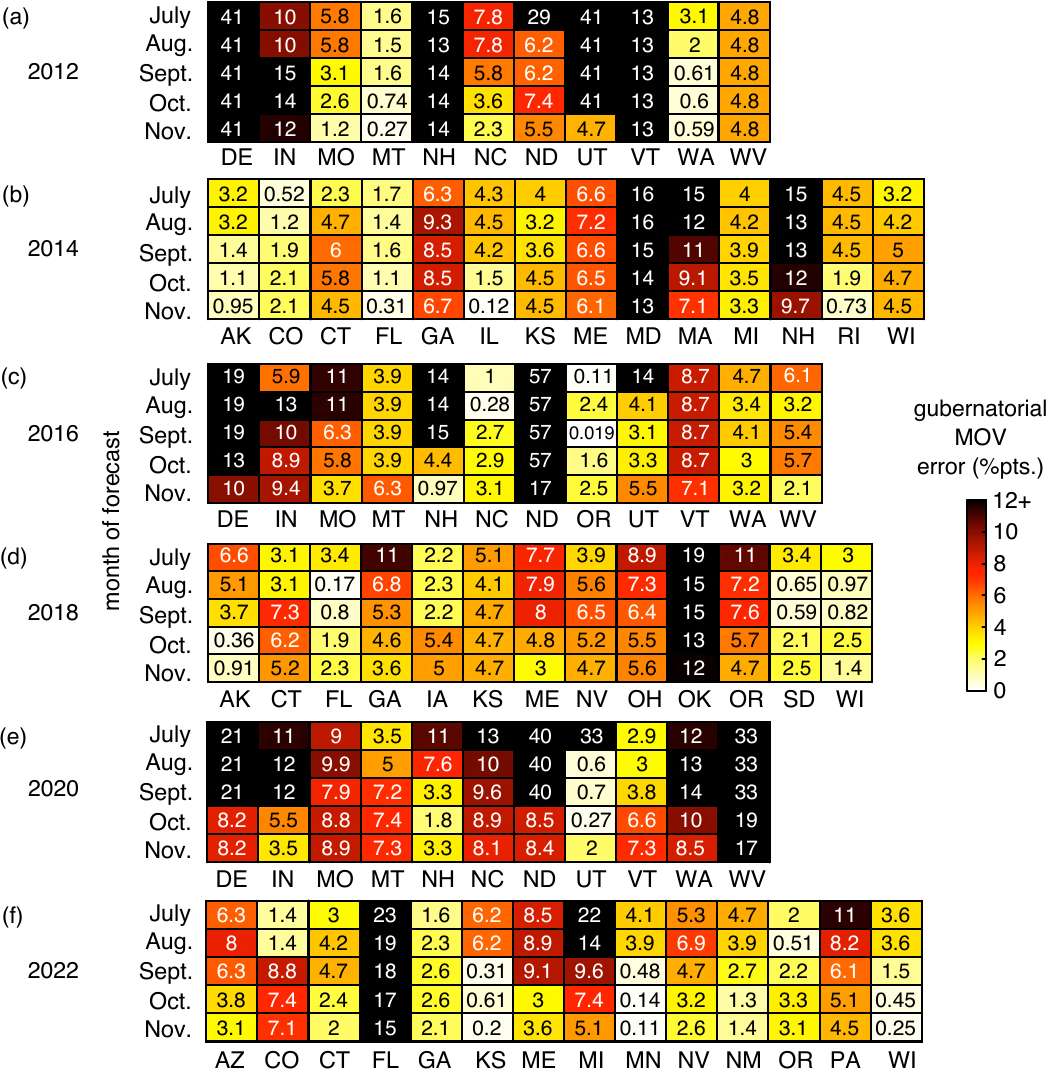}
\caption{MOV errors in time for forecasts of gubernatorial elections in swing states using our baseline approach. We show the MOV errors from July through November for gubernatorial elections in (a) $2012$, (b) $2014$, (c) $2016$, (d) $2018$, (e) $2020$, and (f) $2022$. As is also visible in Fig.~6 in the main manuscript, these results highlight that our MOV error improves as the forecast date nears Election Day. \label{fig:gubMOV}}
\end{figure}

Similarly, in Figures~\ref{fig:gubRatings} and \ref{fig:gubMOV}, we provide the state ratings and MOV errors that we produce using our baseline approach for gubernatorial elections. Our average MOV errors tend to be higher for gubernatorial elections than for senatorial and presidential elections (see Fig.~$6$ in the main manuscript), yet Fig.~\ref{fig:gubRatings} highlights that we incorrectly call between zero and five state gubernatorial races per year for elections between $2012$--$2022$. For $2020$, for example, despite high MOV errors in Fig.~\ref{fig:gubMOV}, we correctly identify the winning gubernatorial candidate by party in every state; see Fig.~\ref{fig:gubRatings}(e). It is also important to note that in several cases polling data were not available from our sources for some state gubernatorial elections. This is the case for Delaware in $2020$, for instance, as is visible by the row of white squares for Delaware in Fig.~\ref{fig:gubRatings}(e) and the MOV error equaling the true MOV for Delaware ($21$ percentage points for the Democratic candidate) in Fig.~\ref{fig:gubMOV}(c).

\subsubsection{Additional sen.\ and gub.\ results using our extended approach}\label{sec:addresults2}

\begin{figure}[t!]
\includegraphics[width=\textwidth]{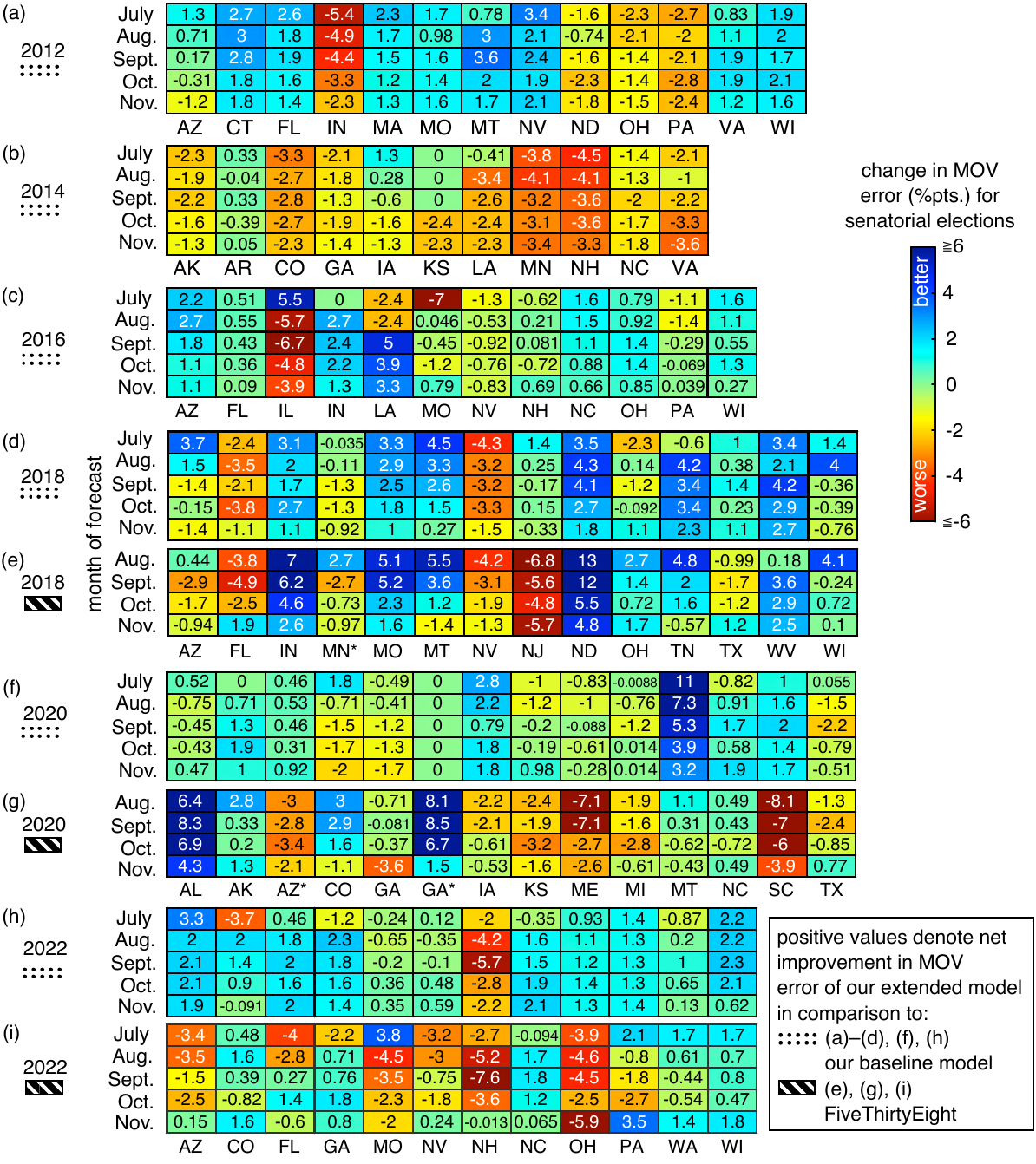}
\vspace{-1.5\baselineskip}
\caption{Comparison of our baseline and extended approaches for senatorial elections by MOV error in swing states. We show the difference in the MOV errors generated by our baseline and extended approaches for (a) $2012$, (b) $2014$, (c) $2016$, (d) $2018$, (f) $2020$, and (h) $2022$. In (e), (g), and (i), we also provide the associated MOV-error differences between our extended approach and FiveThirtyEight (classic version) \cite{FiveThirtyEight2018,FiveThirtyEight2020,FiveThirtyEight2022}. Positive values denote an improvement by our extended approach over the approach in comparison. \label{fig:senMOV2}}
\end{figure}

\begin{figure}[t!]
\includegraphics[width=\textwidth]{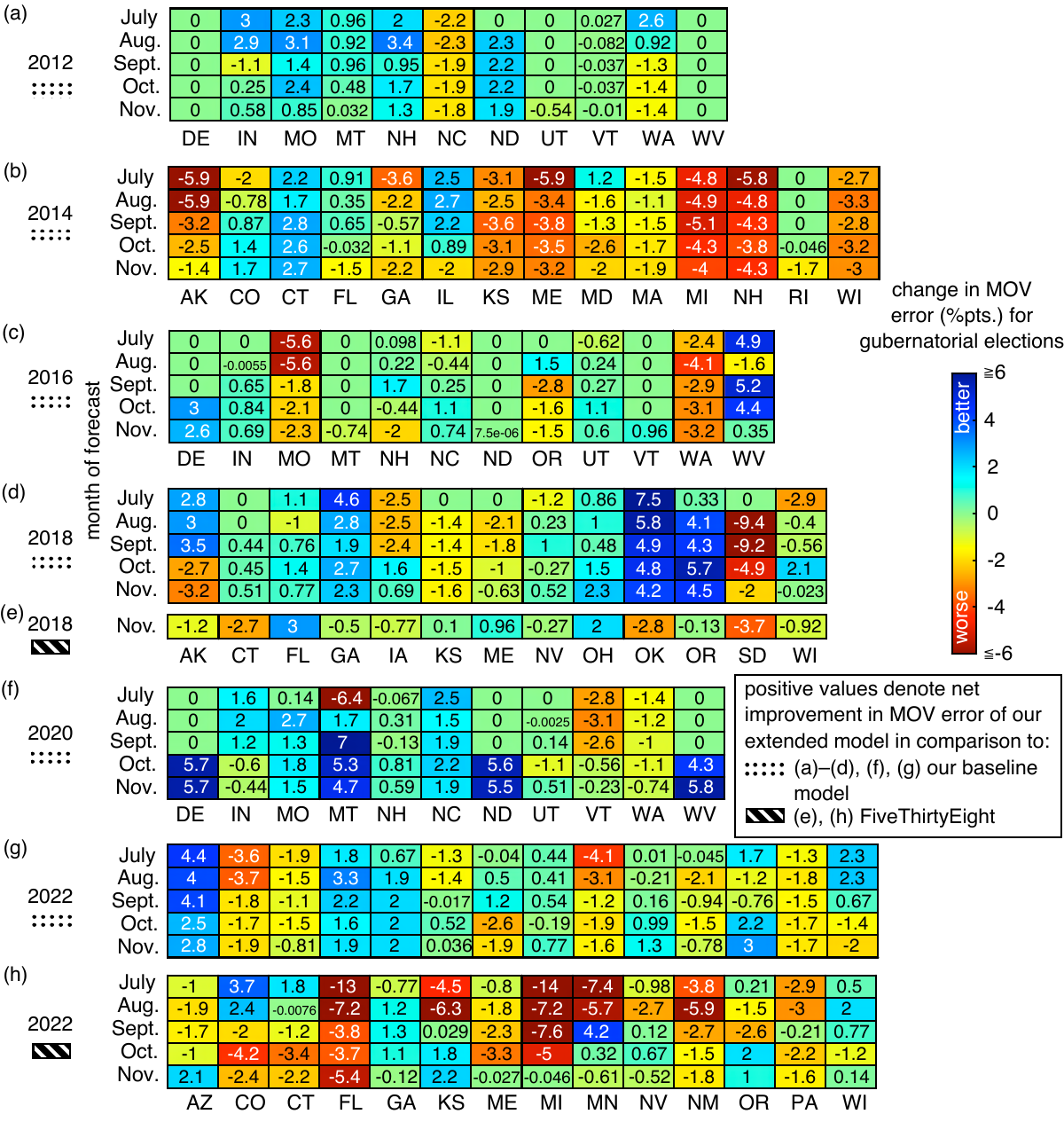}
\vspace{-1\baselineskip}
\caption{Comparison of our baseline and extended approaches for gubernatorial elections by MOV error in swing states. We show the difference in the MOV errors resulting from our baseline and extended approaches for (a) $2012$, (b) $2014$, (c) $2016$, (d) $2018$, (f) $2020$, and (g) $2022$. For context, in (e) and (h), we also present the associated MOV-error differences between our extended approach and FiveThirtyEight (classic version) \cite{FiveThirtyEight2018,FiveThirtyEight2022}. Positive values indicate that our extended approach is better, based on MOV error, than the approach in comparison. Panel (b) highlights that our extended approach performs poorly in $2014$; also see Fig.~\ref{fig:2014}. \label{fig:gubMOV2}}
\end{figure}

\begin{figure}[t!]
\includegraphics[width=\textwidth]{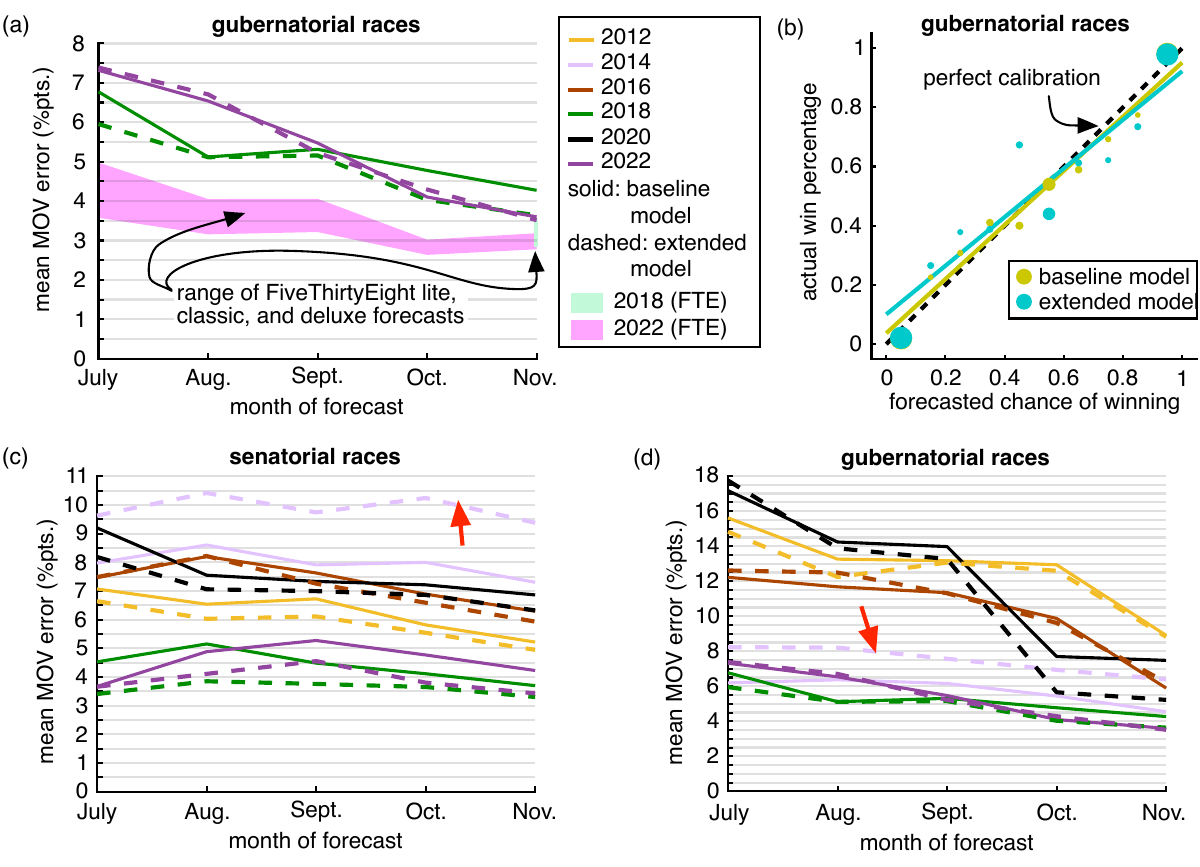}
\vspace{-\baselineskip}
\caption{Summary statistics on the performance of our extended approach for gubernatorial and senatorial elections. (a) Across gubernatorial elections in swing states, the mean MOV error that our extended approach produces is similar to that from our baseline approach. For reference, we show the range of FiveThirtyEight's forecasts from the three versions of their model (lite, classic, and deluxe) \cite{FiveThirtyEight2018,FiveThirtyEight2022} for $2018$ and $2022$. (FiveThirtyEight did not forecast the $2020$ gubernatorial elections.) (b) For gubernatorial elections, our baseline approach appears to be better calibrated than our extended approach. We report calibration using all of our forecasts from July--November for $2012$--$2022$. (c)--(d) Figures~\ref{fig:senMOV2}(b) and \ref{fig:gubMOV2} highlight that our extended approach performs particularly poorly in $2014$. Comparing the mean MOV errors from our baseline (solid) and extended (dashed) approaches in swing states further stresses this point. In (c) senatorial and (d) gubernatorial races, we generally see our extended approach performing similarly or slightly better than our baseline approach, with the significant exception of the $2014$ election (red arrows).\label{fig:2014}}
\end{figure}

Using the MOV error in swing states as our method for measuring forecast performance, we compare our baseline and extended approaches for $2012$--$2022$ senatorial and gubernatorial elections in Figures~\ref{fig:senMOV2} and \ref{fig:gubMOV2}, respectively. Across these two figures, blue-ish tones (i.e., positive numbers) denote net improvement by our extended approach, and red-ish or yellow-ish tones (i.e., negative numbers) indicate that our baseline approach performs better by MOV error in swing states. For completeness, complementing Fig.~9 in the main manuscript, we also provide summary statistics on mean MOV error in time and calibration for gubernatorial forecasts using our two approaches in Fig.~\ref{fig:2014}(a)--(b).

Figure~\ref{fig:senMOV2} shows that our extended approach generally performs better than our baseline approach for senatorial elections. However, the race for the Senate in $2014$ stands out as an exception. We see similar behavior in Fig.~\ref{fig:gubMOV2} for gubernatorial elections, with our baseline approach again out-performing our extended approach for $2014$. This observation that $2014$ displays different behavior than the other election years that we consider is further visible in Fig.~\ref{fig:2014}(c)--(d). For swing states in both $2014$ senatorial and gubernatorial elections, our baseline approach produces monthly forecasts that are about $2$ percentage points closer to the true margin on average, when compared to our extended approach. Figures~\ref{fig:senMOV2}(b) and \ref{fig:gubMOV2}(b) show that our extended approach performs particularly poorly in New Hampshire in $2014$. However, we do correctly identify New Hampshire's elected senator and governor (the Democratic candidate in both cases); see Figures~\ref{fig:senRatings}(b) and \ref{fig:gubRatings}(b). Notably, for the $2014$ senatorial and gubernatorial elections in New Hampshire, our extended approach increases the MOV error because adjusting for pollster history pushes our forecasts overly Democratic. This highlights that past pollster lean does not necessarily translate into current pollster lean.

\subsection{Superstate definitions}\label{sec:superstates}

To reduce the number of equations and parameters in the cRUD model \cite{Volkening2020}, as well as to account for differences in the amount of polling data available for different states, we combine reliably Republican (respectively, Democratic) states or districts into a \textit{Red superstate} (respectively, \textit{Blue superstate}). The state and district races that we forecast as part of the Red or Blue superstates depends on the election type and year; we base these choices on the ratings collected by 270toWin (consensus version) \cite{270toWin}, as well as those reported by FiveThirtyEight \cite{FiveThirtyEight2016,FiveThirtyEight2018,FiveThirtyEight2020,FiveThirtyEight2022}, Sabato's Crystal Ball \cite{Sabato}, \textit{The Huffington Post} \cite{HuffPost2016Method}, Inside Elections \cite{InsideElections}, \textit{The New York Times} \cite{NYT}, and the Cook Political Report \cite{Cook} for some years.

We summarize the state and districts that make up our Red and Blue superstates for each election type and year below. (We use an asterisk to denote special elections.)

\begin{itemize}[noitemsep,nolistsep]
\item \textbf{2004 presidential races:}
\begin{itemize}[noitemsep,nolistsep]
\item \textit{Red superstate} (23 states): Alabama (AL), Alaska (AK), Arizona (AZ), Arkansas (AR), Georgia (GA), Idaho (ID), Indiana (IN), Kansas (KS), Kentucky (KY), Louisiana (LA), Mississippi (MS), Missouri (MO), Montana (MT), Nebraska (NE), North Dakota (ND), Oklahoma (OK), South Carolina (SC), South Dakota (SD), Tennessee (TN), Texas (TX), Utah (UT), West Virginia (WV), and Wyoming (WY);
\item \textit{Blue superstate} (15 races or districts): California (CA), Connecticut (CT), Delaware (DE), District of Columbia (DC), Hawaii (HI), Illinois (IL), Maine (ME), Maryland (MD), Massachusetts (MA), New Jersey (NJ), New York (NY), Oregon (OR), Rhode Island (RI), Vermont (VT), and Washington (WA);
\end{itemize}
\item \textbf{2008 presidential races:} 
\begin{itemize}[noitemsep,nolistsep]
\item \textit{Red superstate} (20 states): AL, AK, AZ, AR, GA, ID, KS, KY, LA, MS, NE, ND, OK, SC, SD, TN, TX, UT, WV, and WY;
\item \textit{Blue superstate} (16 states or districts): CA, CT, DE, DC, HI, IL, ME, MD, MA, NJ, New Mexico (NM), NY, OR, RI, VT, and WA;
\end{itemize}
\item \textbf{2012 presidential races:}
\begin{itemize}[noitemsep,nolistsep]
\item \textit{Red superstate} (23 states): AL, AK, AZ, AR, GA, ID, IN, KS, KY, LA, MS, MO, MT, NE, ND, OK, SC, SD, TN, TX, UT, WV, and WY;
\item \textit{Blue superstate} (16 states or districts): CA, CT, DE, DC, HI, IL, ME, MD, MA, NJ, NM, NY, OR, RI, VT, and WA;
\end{itemize}
\item \textbf{2016 presidential races:}
\begin{itemize}[noitemsep,nolistsep]
\item \textit{Red superstate} (23 states): AL, AK, AZ, AR, GA, ID, IN, KS, KY, LA, MS, MO, MT, NE, ND, OK, SC, SD, TN, TX, UT, WV, and WY;
\item \textit{Blue superstate} (16 states or districts): CA, CT, DE, DC, HI, IL, ME, MD, MA, NJ, NM, NY, OR, RI, VT, and WA;
\end{itemize}
\item \textbf{2020 presidential races:}
\begin{itemize}[noitemsep,nolistsep]
\item \textit{Red superstate} (20 states): AL, AK, AR, ID, IN, KS, KY, LA, MS, MO, MT, NE, ND, OK, SC, SD, TN, UT, WV, and WY;
\item \textit{Blue superstate} (16 states or districts): CA, CT, DE, DC, HI, IL, ME, MD, MA, NJ, NM, NY, OR, RI, VT, and WA;
\end{itemize}
\item \textbf{2024 presidential races:}
\begin{itemize}[noitemsep,nolistsep]
\item \textit{Red superstate} (22 states): AL, AK, AR, ID, IN, IA, KS, KY, LA, MS, MO, MT, NE, ND, OK, SC, SD, TN, TX, UT, WV, and WY;
\item \textit{Blue superstate} (17 states or districts): CA, CO, CT, DE, DC, HI, IL, ME, MD, MA, NJ, NM, NY, OR, RI, VT, and WA;
\end{itemize}
\item \textbf{2012 senatorial races:}
\begin{itemize}[noitemsep,nolistsep]
\item \textit{Red superstate} (6 states): MS, NE, TN, TX, UT, and WY;
\item \textit{Blue superstate} (14 states): CA, DE, HI, ME, MD, Michigan (MI), Minnesota (MN), NJ, NM, NY, RI, VT, WA, and WV;
\end{itemize}
\item \textbf{2014 senatorial races:}
\begin{itemize}[noitemsep,nolistsep]
\item \textit{Red superstate} (16 states): AL, ID, KY, ME, MS, MT, NE, OK, OK*, SC, SC*, SD, TN, TX, WV, and WY;
\item \textit{Blue superstate} (9 states): DE, HI*, IL, MA, MI, NJ, NM, OR, and RI;
\end{itemize}
\item \textbf{2016 senatorial races:}
\begin{itemize}[noitemsep,nolistsep]
\item \textit{Red superstate} (13 states): AL, AK, AR, GA, ID, IA, KS, KY, ND, OK, SC, SD, and UT;
\item \textit{Blue superstate} (8 states): Colorado (CO), CT, HI, MD, NY, OR, VT, and WA;
\end{itemize}
\item \textbf{2018 senatorial races:}
\begin{itemize}[noitemsep,nolistsep]
\item \textit{Red superstate} (5 states): MS, MS*, NE, UT, and WY;
\item \textit{Blue superstate} (15 states): CT, DE, HI, ME, MD, MA, MI, MN, NM, NY, Pennsylvania (PA), RI, VT, VA, and WA;
\end{itemize}
\item \textbf{2020 senatorial races:}
\begin{itemize}[noitemsep,nolistsep]
\item \textit{Red superstate} (11 states): AR, ID, KY, LA, MS, NE, OK, SD, TN, WV, and WY;
\item \textit{Blue superstate} (10 states): DE, IL, MA, MN, NH, NJ, NM, OR, RI, and VA;
\end{itemize}
\item \textbf{2022 senatorial races:}
\begin{itemize}[noitemsep,nolistsep]
\item \textit{Red superstate} (15 states): AL, AK, AR, ID, IN, Iowa (IA), KS, KY, LA, ND, OK, OK*, SC, SD, and UT;
\item \textit{Blue superstate} (8 states): CA, CT, HI, IL, MD, NY, OR, and VT;
\end{itemize}
\item \textbf{2024 senatorial races:}
\begin{itemize}[noitemsep,nolistsep]
\item \textit{Red superstate} (10 states): IN, MS, MO, NE, NE*, ND, TN, UT, WV, WY;
\item \textit{Blue superstate} (15 states): CA, CT, DE, HI, ME, MD, MA, MN, NJ, NM, NY, RI, VT, VA, WA;
\end{itemize}
\item \textbf{2012 gubernatorial races:}
\begin{itemize}[noitemsep,nolistsep]
\item No superstates;
\end{itemize}
\item \textbf{2014 gubernatorial races:}
\begin{itemize}[noitemsep,nolistsep]
\item \textit{Red superstate} (15 states): AL, AZ, AR, ID, IA, NE, Nevada (NV), NM, OH, OK, SC, SD, TN, TX, and WY;
\item \textit{Blue superstate} (7 states): CA, HI, MN, NY, OR, PA, and VT;
\end{itemize}
\item \textbf{2016 gubernatorial races:}
\begin{itemize}[noitemsep,nolistsep]
\item No superstates;
\end{itemize}
\item \textbf{2018 gubernatorial races:}
\begin{itemize}[noitemsep,nolistsep]
\item \textit{Red superstate} (13 states): AL, AZ, AR, ID, MD, MA, NE, New Hampshire (NH), SC, TN, TX, VT, and WY;
\item \textit{Blue superstate} (10 states): CA, CO, HI, IL, MI, MN, NM, NY, PA, and RI;
\end{itemize}
\item \textbf{2020 gubernatorial races:}
\begin{itemize}[noitemsep,nolistsep]
\item No superstates;
\end{itemize}
\item \textbf{2022 gubernatorial races:}
\begin{itemize}[noitemsep,nolistsep]
\item \textit{Red superstate} (15 states): AL, AK, AR, ID, IA, NE, NH, OH, OK, SC, SD, TN, TX, VT, and WY;
\item \textit{Blue superstate} (7 states): CA, HI, IL, MD, MA, NY, and RI.
\end{itemize}
\item \textbf{2024 gubernatorial races:}
\begin{itemize}[noitemsep,nolistsep]
\item No superstates;
\end{itemize}
\end{itemize}
Because there were not many states holding gubernatorial elections in 2012, 2016, 2020, and 2024, we do not use superstates for these years.

\section*{Acknowledgments}
A.V. thanks Daniel F. Linder, Mason A. Porter, and Grzegorz A. Rempala for their helpful comments, encouragement, and friendship. Our parameter-estimation programs build on Daniel's software \cite{Gitlab_elections}. We are also grateful to David Leip (of \textit{Dave Leip's Atlas of U.S. Presidential Elections}) \cite{Leip} for aggregating election results and permitting us to use these data to quantify accuracy. We thank the FiveThirtyEight \cite{538mainNew,538mainOld}, RealClearPolitics \cite{RealClearPolling}, and HuffPost Pollster \cite{HuffPostPollster,HuffPostAPI} teams for aggregating polls.

\section*{Funding}
The work of R.B., J.C., and T.P. has been supported in part by the EURO Summer Undergraduate Research Fellowship (SURF) program and by the Office of Undergraduate Research (OUR) Scholars program at Purdue University. The work of M.L. has been supported in part by the EURO SURF program at Purdue University. The work of A.R. has been supported in part by the OUR Scholars program at Purdue University. The work of W.L.H. and E.M. has been supported in part by the Undergraduate Research Assistant (URA) program at Northwestern University. The work of C.M.L. has been supported in part by the Northwestern University URA program and by the National Science Foundation (NSF) under grant no.~$1547394$. The work of A.V. has been supported in part by the NSF under grant no. DMS-$1764421$ and by the Simons Foundation/SFARI under grant no.~$597491$-RWC.

\bibliographystyle{siamplain}
\bibliography{electionBib}

\begin{thebibliography}{100}

\bibitem{FiveThirtyEightGitHub}
{\em {fivethirtyeight/data: Data and code behind the articles and graphics at
  FiveThirtyEight}}.
\newblock \url{https://github.com/fivethirtyeight/data/tree/master}.
\newblock Last accessed: 22 May 2024.

\bibitem{Wayback}
{\em {Internet archive: Wayback Machine}}.
\newblock \url{https://web.archive.org}.
\newblock {Last accessed: 30 Oct 2024}.

\bibitem{RealClearPolitics2004}
{\em {RealClearPolitics Polls: 2004}}.
\newblock
  \url{https://www.realclearpolitics.com/Presidential_04/ma_polls.html}, 2004.
\newblock Accessed for data: 5 Apr 2020 and 5 May 2023.

\bibitem{FedRegister2004}
{\em {Estimates of the Voting Age Population for 2004. A Notice by the Commerce
  Department on 01/18/2005. Agency: Office of the Secretary, Commerce}},
  Federal Register Notices, 70 (2005).

\bibitem{RealClearPolitics2008}
{\em {RealClearPolitics Polls: 2008}}.
\newblock
  \url{https://www.realclearpolitics.com/epolls/2008/president/co/colorado_mccain_vs_obama-546.html#polls},
  2008.
\newblock Accessed for data: 4 Apr 2020 and 5 May 2023.

\bibitem{FedRegister2008}
{\em {Estimates of the Voting Age Population for 2008. A Notice by the Commerce
  Department on 01/27/2009. Agency: Office of the Secretary, Commerce}},
  Federal Register Notices, 74 (2009).

\bibitem{FedRegister2012}
{\em {Estimates of the Voting Age Population for 2012. A Notice by the Commerce
  Department on 01/30/2013. Agency: Office of the Secretary, Commerce}},
  Federal Register Notices, 78 (2013).

\bibitem{FedRegister2014}
{\em {Estimates of the Voting Age Population for 2014. A Notice by the Commerce
  Department on 02/06/2015. Agency: Office of the Secretary, Commerce}},
  Federal Register Notices, 80 (2015).

\bibitem{FedRegister2016}
{\em {Estimates of the Voting Age Population for 2016. A Notice by the Commerce
  Department on 01/30/2017. Agency: Office of the Secretary, Commerce}},
  Federal Register Notices, 82 (2017).

\bibitem{FedRegister20202018}
{\em {Estimates of the Voting Age Population for 2018. A Notice by the Commerce
  Department on 10/04/2019. Agency: Office of the Secretary, Commerce}},
  Federal Register Notices, 84 (2019).

\bibitem{FedRegister2022}
{\em {Estimates of the Voting Age Population for 2020. A Notice by the Commerce
  Department on 05/06/2021. Agency: Office of the Secretary, Commerce}},
  Federal Register Notices, 86 (2021).

\bibitem{HuffPostPollster}
{\em {HuffPost Pollster}}.
\newblock \url{https://elections.huffingtonpost.com/pollster}, (2022.
\newblock Last accessed: 16 Oct 2022.

\bibitem{HuffPostAPI}
{\em {HuffPost Pollster API v2}}.
\newblock \url{https://elections.huffingtonpost.com/pollster/api/v2}, 2022.
\newblock Last accessed: 16 Oct 2022.

\bibitem{270toWin}
{\em {270toWin}}.
\newblock \url{https://www.270towin.com}, 2024.
\newblock {Last accessed: 21 May 2024}.

\bibitem{538mainNew}
{\em {538: abc NEWS}}.
\newblock \url{https://abcnews.go.com/538}, 2024.
\newblock Last accessed: 8 May 2024.

\bibitem{538mainOld}
{\em {FiveThirtyEight: All Posts Tagged ``Forecasts" (archived version of
  site)}}.
\newblock \url{https://fivethirtyeight.com/tag/forecasts/}, 2024.
\newblock Last accessed: 8 May 2024.

\bibitem{RealClearPolling}
{\em {RealClearPolling: 2024 RCP Electoral College Map}}.
\newblock
  \url{https://www.realclearpolling.com/maps/president/2024/toss-up/electoral-college},
  2024.
\newblock Last accessed: 22 Oct 2024.

\bibitem{RealClearPollingData}
{\em {RealClearPolling: Latest Polls}}.
\newblock \url{https://www.realclearpolling.com/latest-polls/election}, 2024.
\newblock Last accessed: 22 May 2024.

\bibitem{Cook}
{\em {The Cook Political Report: Ratings}}.
\newblock \url{https://www.cookpolitical.com/ratings}, 2024.
\newblock Last accessed: 22 May 2024.

\bibitem{Abramowitz1988}
{\sc A.~I. Abramowitz}, {\em An improved model for predicting presidential
  election outcomes}, PS Political Sci Politics, 21 (1988), pp.~843--847.

\bibitem{Abramowitz2008}
{\sc A.~I. Abramowitz}, {\em {Forecasting the 2008 Presidential Election with
  the Time-for-Change Model}}, PS Political Sci Politics, 41 (2008),
  pp.~691--695.

\bibitem{turnout}
{\sc E.~Alabrese}, {\em {National Polls, Local Preferences and Voters’
  Behaviour: Evidence from the UK General Elections}}.
\newblock {Quantitative and Analytical Political Economy Research Centre,
  Warwick Economics Research Paper Series,
  \url{https://warwick.ac.uk/fac/soc/economics/research/centres/qapec/discussionpapers}},
  2022.

\bibitem{NYT}
{\sc C.~Baker, L.~B. Jensen, D.~C. Ademola~Bello, N.~Cohn, M.~C. Escobar,
  A.~Daniel, R.~Igielnik, K.~R. Lai, J.~C. Lee, A.~Lemonides, A.~Sun,
  R.~Taylor, I.~White, K.~Bayrakdarian, A.~Elkeurti, A.~Fischer, A.~Park,
  J.~Patel, D.~Simmons-Ritchie, E.~Singer, and J.~Thomas}, {\em {Election 2024
  Polls: Harris vs. Trump}}.
\newblock
  \url{https://www.nytimes.com/interactive/2024/us/elections/polls-president.html},
  2024.
\newblock Last accessed: 22 Oct 2024.

\bibitem{FiveThirtyEightLatestPolls}
{\sc R.~Best, A.~Bycoffe, C.~Groskopf, R.~King, E.~Koeze, D.~Mehta, J.~Mithani,
  M.~Radcliffe, A.~Wiederkehr, J.~Wolfe, A.~Jones-Rooy, N.~Rakich, D.~Shan,
  S.~Frostenson, J.~Mason, A.~Mangan, and C.~Yee}, {\em {FiveThirtyEight:
  Latest Polls}}.
\newblock \url{https://projects.fivethirtyeight.com/polls/}, 2022.
\newblock 2018 polls last accessed: 2 July 2021. 2020 polls last accessed: 11
  Apr 2021. 2022 polls last accessed: 5 Nov 2022. 2024 polls last accessed Oct
  2024.

\bibitem{IdeaSpread}
{\sc L.~M.~A. Bettencourt, A.~Cintr\'{o}n-Arias, D.~I. Kaiser, and
  C.~Castillo-Chav\'{e}z}, {\em The power of a good idea: {Q}uantitative
  modeling of the spread of ideas from epidemiological models}, Phys A, 364
  (2006), pp.~513--536.

\bibitem{Biondo2018}
{\sc A.~E. Biondo, A.~Pluchino, and A.~Rapisarda}, {\em Modeling surveys
  effects in political competitions}, Phys A: Stat Mech Appl, 503 (2018),
  pp.~714--726.

\bibitem{FiveThirtyEightBrier}
{\sc J.~Boice, G.~Wezerek, L.~Bronner, and A.~Bycoffe}, {\em {How Good Are
  FiveThirtyEight Forecasts?}}
\newblock \url{https://projects.fivethirtyeight.com/checking-our-work/}, 2023.
\newblock {Last accessed: 8 May 2024}.

\bibitem{FrenchRiots}
{\sc L.~Bonnasse-Gahot, H.~Berestycki, M.-A. Depuiset, M.~B. Gordon,
  S.~Roch\'{e}, N.~Rodriguez, and J.-P. Nadal}, {\em Epidemiological modelling
  of the 2005 {F}rench riots: {a} spreading wave and the role of contagion},
  Sci Rep, 8 (2018).

\bibitem{BottcherPLOS}
{\sc L.~Bottcher, H.~J. Herrmann, and H.~Gersbach}, {\em Clout, activists and
  budget: {T}he road to presidency}, PLOS ONE, 13 (2018), p.~e0193199.

\bibitem{Braha}
{\sc D.~Braha and M.~A.~M. de~Aguiar}, {\em Voting contagion: {M}odeling and
  analysis of a century of {U.S.} presidential elections}, PLOS ONE, 12 (2017),
  p.~e0177970.

\bibitem{Gitlab_elections_v2}
{\sc R.~Branstetter, S.~Chian, J.~Cromp, W.~L. He, C.~M. Lee, M.~Liu,
  E.~Mansell, M.~Paranjape, T.~Pattanashetty, A.~Rodrigues, and A.~Volkening},
  {\em {Code associated with ``How time and pollster history affect U.S.
  election forecasts under a compartmental modeling approach"}}.
\newblock
  \url{https://gitlab.com/alexandriavolkening/forecasting-elections-using-compartmental-models-2},
  2024.

\bibitem{2022forecasts}
{\sc R.~Branstetter, M.~Liu, M.~Paranjape, and A.~Volkening}, {\em {CRUD: A
  compartmental Republican--Undecided--Democratic model for forecasting U.S.
  elections}}.
\newblock \url{https://c-r-u-d.gitlab.io/2022/}, 2022.

\bibitem{ccc}
{\sc R.~Brauer and C.~Castillo-Chavez}, {\em {Mathematical Models in Population
  Biology and Epidemiology}}, Springer-Verlag, Heidelberg, Germany, 2nd~ed.,
  2012.

\bibitem{Brier}
{\sc G.~W. Brier}, {\em Verification of forecasts expressed in terms of
  probability}, Mon Weather Rev, 78 (1950), pp.~1--3.

\bibitem{FTEratingsWork}
{\sc A.~Bycoffe, M.~Radcliffe, C.~Burton, D.~Mehta, A.~Mangan, and N.~Silver},
  {\em {FiveThirtyEight: How our pollster ratings work}}.
\newblock \url{https://projects.fivethirtyeight.com/pollster-ratings/}, 2023.
\newblock Last accessed: 5 May 2023. Now available through WayBack Machine.

\bibitem{byrd1995limited}
{\sc R.~H. Byrd, P.~Lu, J.~Nocedal, and C.~Zhu}, {\em {A Limited Memory
  Algorithm for Bound Constrained Optimization}}, SIAM J Sci Comput, 16 (1995),
  pp.~1190--1208.

\bibitem{Campbell1992}
{\sc J.~E. Campbell}, {\em Forecasting the presidential vote in the states}, Am
  J Pol Sci, 36 (1992), pp.~386--407.

\bibitem{Campbell}
{\sc J.~E. Campbell}, {\em {Polls and Votes: The Trial-Heat Presidential
  Election Forecasting Model, Certainty, and Political Campaigns}}, Am Politics
  Res, 24 (1996), pp.~408--433.

\bibitem{Campbell2014}
{\sc J.~E. Campbell}, {\em Issues in presidential election forecasting:
  Election margins, incumbency, and model credibility}, PS: Political Sci
  Politics, 47 (2014), pp.~301--303.

\bibitem{castellano09}
{\sc C.~Castellano, S.~Fortunato, and V.~Loreto}, {\em Statistical physics of
  social dynamics}, Rev Mod Phys, 81 (2009), pp.~591--646.

\bibitem{Chen2023}
{\sc Y.~Chen, R.~Garnett, and J.~M. Montgomery}, {\em Polls, context, and time:
  A dynamic hierarchical bayesian forecasting model for us senate elections},
  Political Anal, 31 (2023), pp.~113--133.

\bibitem{2020forecasts}
{\sc S.~Chian, W.~L. He, C.~M. Lee, D.~F. Linder, M.~A. Porter, G.~A. Rempala,
  and A.~Volkening}, {\em {2020 U.S. Election Forecasts with a Compartmental
  Model}}.
\newblock \url{https://modelingelectiondynamics.gitlab.io/2020-forecasts/},
  2020.

\bibitem{Cook2014}
{\sc C.~E. Cook and D.~Wasserman}, {\em Recalibrating ratings for a new
  normal}, PS: Political Sci Politics, 47 (2014), pp.~304--308.

\bibitem{2024forecasts}
{\sc J.~Cromp, T.~Pattanashetty, A.~Rodrigues, and A.~Volkening}, {\em {CRUD
  2024: A compartmental Republican--Undecided--Democratic model for forecasting
  U.S. elections}}.
\newblock \url{https://c-r-u-d.gitlab.io/2024/}, 2024.

\bibitem{deStefano2022}
{\sc D.~De~Stefano, F.~Pauli, and N.~Torelli}, {\em Preelectoral polls
  variability: {A} hierarchical {B}ayesian model to assess the role of house
  effects with application to {I}talian elections}, Ann Appl Stat, 16 (2022).

\bibitem{Diekmann}
{\sc O.~Diekmann and J.~A.~P. Heesterbeek}, {\em {Mathematical Epidemiology of
  Infectious Diseases: Model Building, Analysis and Interpretation}}, Wiley,
  New York, 2000.

\bibitem{Erikson1996}
{\sc R.~S. Erikson and C.~Wlezien}, {\em Of time and presidential election
  forecasts}, PS: Political Sci Politics, 29 (1996), pp.~37--39.

\bibitem{Fernandez2014}
{\sc J.~Fern\'{a}ndez-Gracia, K.~Suchecki, J.~J. Ramasco, M.~San~Miguel, and
  V.~M. Eguiluz}, {\em {Is the Voter Model a Model for Voters?}}, Phys Rev
  Lett, 112 (2014), p.~158701.

\bibitem{Bigten}
{\sc {FiveThirtyEight}}, {\em {GitHub: FiveThirtyEight Pollster Ratings: Raw
  Polls}}.
\newblock
  \url{https://raw.githubusercontent.com/fivethirtyeight/data/master/pollster-ratings/raw-polls.csv}.
\newblock {Last accessed: 5 May 2023. No longer available; see WayBack
  Machine}.

\bibitem{Forsberg2015}
{\sc O.~J. Forsberg and M.~E. Payton}, {\em Analysis of battleground state
  presidential polling performances, 2004-2012}, Stat Public Policy, 2 (2015),
  pp.~1--10.

\bibitem{EducationData2022}
{\sc {FRED Economic Data, Federal Reserve Bank of St. Louis}}, {\em
  {Educational Attainment, Annual: Bachelor's Degree or Higher by State
  (2020)}}.
\newblock
  \url{https://fred.stlouisfed.org/release/tables?rid=330&eid=391444&od=2020-01-01#}.
\newblock {Last accessed: 24 July 2022}.

\bibitem{EducationData2024}
{\sc {FRED Economic Data, Federal Reserve Bank of St. Louis}}, {\em
  {Educational Attainment, Annual: Bachelor's Degree or Higher by State
  (2022)}}.
\newblock \url{https://fred.stlouisfed.org/release/tables?eid=391444&rid=330}.
\newblock {Last accessed: 2 Feb. 2024}.

\bibitem{GalamTrump}
{\sc S.~Galam}, {\em The {T}rump phenomenon: An explanation from sociophysics},
  Int J Mod Phys B, 31 (2017), p.~1742015.

\bibitem{Gelman2021}
{\sc A.~Gelman}, {\em Failure and success in political polling and election
  forecasting}, Stat Public Policy, 8 (2021), pp.~67--72.

\bibitem{Gelman2020}
{\sc A.~Gelman, J.~Hullman, C.~Wlezien, and G.~E. Morris}, {\em Information,
  incentives, and goals in election forecasts}, Judgm Decis Mak, 15 (2020),
  pp.~863--880.

\bibitem{Gelman}
{\sc A.~Gelman and G.~King}, {\em {Why are American Presidential Election
  Campaign Polls So Variable When Votes Are So Predictable?}}, Br J Political
  Sci, 23 (1993), pp.~409--451.

\bibitem{InsideElections}
{\sc N.~L. Gonzales, J.~Rubashkin, B.~Wascher, and S.~Rothenberg}, {\em {Inside
  Elections with Nathan L. Gonzales, Nonpartisan Analysis}}.
\newblock \url{https://insideelections.com}, 2024.
\newblock Last accessed: 22 May 2024.

\bibitem{Heidemanns2020}
{\sc M.~Heidemanns, A.~Gelman, and G.~E. Morris}, {\em {An updated dynamic
  Bayesian forecasting model for the US presidential election}}, Harvard Data
  Science Review,  (2020).

\bibitem{Hernandez2018}
{\sc A.~R. Hern\'{a}ndez, C.~Gracia-L\'{a}zaro, E.~Brigatti, and Y.~Moreno},
  {\em A networked voting rule for democratic representation}, Royal Soc Open
  Sci, 5 (2018), p.~172265.

\bibitem{HethcoteReview}
{\sc H.~W. Hethcote}, {\em {The Mathematics of Infectious Diseases}}, SIAM Rev,
  42 (2000), pp.~599--653.

\bibitem{Higham}
{\sc D.~J. Higham}, {\em An algorithmic introduction to numerical simulation of
  stochastic differential equations}, SIAM Rev, 43 (2001), pp.~525--546.

\bibitem{Jackman}
{\sc S.~Jackman}, {\em Pooling the polls over an election campaign}, Aust J
  Political Sci, 40 (2005), pp.~499--517.

\bibitem{Jackman2014}
{\sc S.~Jackman}, {\em The predictive power of uniform swing}, PS: Political
  Sci Politics, 47 (2014), pp.~317--321.

\bibitem{HuffPost2016Method}
{\sc N.~Jackson and A.~Hooper}, {\em {Huffington Post Election 2016 Forecast:
  President}}.
\newblock \url{http://elections.huffingtonpost.com/2016/forecast/president},
  2016.
\newblock Accessed: 22 May 2024.

\bibitem{AAPOR}
{\sc C.~Kennedy, M.~Blumenthal, S.~Clement, J.~D. Clinton, C.~Durand,
  C.~Franklin, K.~McGeeney, L.~Miringoff, K.~Olson, D.~Rivers, L.~Saad, G.~E.
  Witt, and C.~Wlezien}, {\em An evaluation of the 2016 election polls in the
  {United States}}, Public Opin Q, 82 (2018), pp.~1--33.

\bibitem{Kermack700}
{\sc W.~O. Kermack and A.~G. McKendrick}, {\em A contribution to the
  mathematical theory of epidemics}, Proc R Soc London, 115 (1927),
  pp.~700--721.

\bibitem{Leip}
{\sc D.~Leip}, {\em {Dave Leip's Atlas of U.S. Presidential Elections}}.
\newblock \url{https://uselectionatlas.org}, 2024.
\newblock {Last accessed: 21 May 2024. 2022 margins accessed on 24 Oct 2023.
  2004--2020 margins accessed from 1--6 July 2021.}

\bibitem{Lewis-Beck1984}
{\sc M.~S. Lewis-Beck and T.~W. Rice}, {\em Forecasting presidential elections:
  A comparison of naive models}, Political Behav, 6 (1984), pp.~9--21.

\bibitem{Linzer}
{\sc D.~A. Linzer}, {\em {Dynamic Bayesian Forecasting of Presidential
  Elections in the States}}, J Am Stat Assoc, 108 (2013), pp.~124--134.

\bibitem{Marvel2012}
{\sc S.~A. Marvel, H.~Hong, A.~Papush, and S.~H. Strogatz}, {\em Encouraging
  moderation: clues from a simple model of ideological conflict}, Phys Rev
  Lett, 109 (2012), p.~118702.

\bibitem{FiveThirtyEightRatings}
{\sc E.~Mej\'{i}a, A.~Bycoffe, M.~Radcliffe, C.~Burton, D.~Mehta, A.~Mangan,
  and C.~Groskopf}, {\em {FiveThirtyEight: FiveThirtyEight's Pollster
  Ratings}}.
\newblock \url{https://projects.fivethirtyeight.com/pollster-ratings/}, 2023.
\newblock Last accessed: 2 May 2023.

\bibitem{PANA2009}
{\sc C.~Panagopoulos}, {\em Polls and elections: Preelection poll accuracy in
  the 2008 general elections}, Pres Stud Q, 39 (2009), pp.~896--907.

\bibitem{Pasek2015}
{\sc J.~Pasek}, {\em The polls–review: Predicting elections: Considering
  tools to pool the polls}, Public Opin Q, 79 (2015), pp.~594--619.

\bibitem{Pickup2008}
{\sc M.~Pickup and R.~Johnston}, {\em Campaign trial heats as election
  forecasts: Measurement error and bias in 2004 presidential campaign polls},
  Int J Forecast, 24 (2008), pp.~272--284.

\bibitem{PorterGleeson}
{\sc M.~A. Porter and J.~P. Gleeson}, {\em Dynamical Systems on Networks: A
  Tutorial}, vol.~4 of Front Appl Dyn Syst Rev Tutor, Springer International
  Publishing, Cham, Switzerland, 2016.

\bibitem{prosser_mellon_2018}
{\sc C.~Prosser and J.~Mellon}, {\em {The Twilight of the Polls? A Review of
  Trends in Polling Accuracy and the Causes of Polling Misses}}, Gov Oppos, 53
  (2018), pp.~757--709.

\bibitem{R}
{\sc {R Core Team}}, {\em R: A Language and Environment for Statistical
  Computing}, R Foundation for Statistical Computing, Vienna, Austria, 2018,
  \url{https://www.R-project.org/}.

\bibitem{Restrepo2009}
{\sc J.~M. Restrepo, R.~C. Rael, and J.~M. Hyman}, {\em Modeling the influence
  of polls on elections: a population dynamics approach}, Public Choice, 140
  (2009), pp.~395--420.

\bibitem{Sabato}
{\sc L.~J. Sabato, K.~Kondik, and J.~M. Coleman.}, {\em {Sabato's Crystal
  Ball}}.
\newblock \url{https://centerforpolitics.org/crystalball/}, The Center for
  Politics at the University of Virginia, 2024.
\newblock Last accessed: 22 May 2024.

\bibitem{Miller2020}
{\sc D.~Sabin-Miller and D.~M. Abrams}, {\em {When pull turns to shove: A
  continuous-time model for opinion dynamics}}, Phys Rev Research, 2 (2020),
  p.~043001.

\bibitem{Selb2023}
{\sc P.~Selb, S.~Chen, J.~K\"{o}rtner, and P.~Bosch}, {\em Bias and variance in
  multiparty election polls}, Public Opin Q, 87 (2023), pp.~1025--1037.

\bibitem{Selb2016}
{\sc P.~Selb and S.~Munzert}, {\em Forecasting the 2013 german bundestag
  election using many polls and historical election results}, German Politics,
  25 (2016), pp.~73--83.

\bibitem{Shirani-Mehr2018}
{\sc H.~Shirani-Mehr, D.~Rothschild, S.~Goel, and A.~Gelman}, {\em
  Disentangling bias and variance in election polls}, J Am Stat Assoc, 113
  (2018), pp.~607--614.

\bibitem{SilverBook}
{\sc N.~Silver}, {\em The Signal and the Noise: {W}hy So Many Predictions Fail
  --- But Some Don't}, Penguin Press, New York City, NY, USA, 2012.

\bibitem{SilverBulletinAnnoucement}
{\sc N.~Silver}, {\em {Announcing 2024 election model plans}}.
\newblock
  \url{https://www.natesilver.net/p/announcing-2024-election-model-plans},
  2024.
\newblock Last accessed: 8 May 2024.

\bibitem{SilverBulletin}
{\sc N.~Silver}, {\em {Silver Bulletin}}.
\newblock \url{https://www.natesilver.net}, 2024.
\newblock Last accessed: 8 May 2024.

\bibitem{FiveThirtyEight2018}
{\sc N.~Silver, J.~Boice, E.~Brillhart, A.~Bycoffe, R.~Dottle, L.~Eastridge,
  R.~King, E.~Koeze, A.~Scheinkman, G.~Wezerek, J.~Wolfe, D.~Dienhart,
  A.~Jones-Rooy, D.~Mehta, M.~Nguyen, N.~Rakich, D.~Shan, and G.~Skelley}, {\em
  Election 2018: Fivethirtyeight}.
\newblock
  \url{https://projects.fivethirtyeight.com/2018-midterm-election-forecast/house/},
  2018.
\newblock {Last accessed: 22 May 2024. 2018 forecasts accessed: 2 July 2021}.

\bibitem{FiveThirtyEight2020}
{\sc N.~Silver, C.~Groskopf, R.~Best, J.~Mithani, A.~Wiederkehr, M.~Cohen,
  S.~Frostenson, E.~Scherer, F.~Buonocore, J.~Ellis, J.~Boice, A.~Bycoffe,
  E.~Mej\'{i}a, J.~Wolfe, Y.~Yuan, C.~Barry, J.~Mason, D.~Mhta, M.~Radcliffe,
  and D.~Shan}, {\em {FiveThirtyEight 2020}}.
\newblock \url{https://projects.fivethirtyeight.com/2020-election-forecast/},
  2020.
\newblock {Last accessed: 21 May 2024. 2020 forecasts accessed: 2 July 2021}.

\bibitem{FiveThirtyEight2022}
{\sc N.~Silver, C.~Groskopf, S.~Frostenson, M.~Ganesan, J.~Mason, R.~Best,
  E.~Mej\'{i}a, J.~Boice, A.~Bycoffe, E.~Scherer, F.~Buonocore, J.~Ellis,
  C.~Burton, M.~Radcliffe, C.~Barry, M.~Cohen, D.~Mehta, J.~Mithani, D.~Shan,
  A.~Wiederkehr, J.~Wolfe, and Y.~Yuan}, {\em {FiveThirtyEight 2022}}.
\newblock \url{https://projects.fivethirtyeight.com/2022-election-forecast/},
  2022.
\newblock {Last accessed: 21 May 2024. 2022 forecasts accessed: 27 Sept 2023}.

\bibitem{FiveThirtyEight2016}
{\sc N.~Silver, J.~Kanjana, D.~Mehta, J.~Boice, A.~Bycoffe, M.~Conlen,
  R.~Fischer-Baum, R.~King, E.~Koeze, A.~McCann, A.~Scheinkman, and
  G.~Wezerek}, {\em {FiveThirtyEight: 2016 Election Forecast}}.
\newblock \url{https://projects.fivethirtyeight.com/2016-election-forecast/},
  2016.
\newblock Last accessed: 22 May 2024.

\bibitem{TheEconomist}
{\sc {The Economist with Andrew Gelman and colleagues}}, {\em {The Economist:
  US election 2024}}.
\newblock
  \url{https://www.economist.com/interactive/us-2024-election/prediction-model/president},
  2024.
\newblock Last accessed: 22 Oct 2024.

\bibitem{Traugott2014}
{\sc M.~W. Traugott}, {\em Public opinion polls and election forecasting}, PS:
  Political Sci Politics, 47 (2014), pp.~342--344.

\bibitem{EducationData2020201820162014201220082004}
{\sc {U.S. Census Bureau}}, {\em {2018 American Community Survey 1-Year
  Estimates: Table R1502: Percent of People 25 Years and Older Who Have
  Completed a Bachelor's Degree}}.
\newblock
  \url{https://www2.census.gov/programs-surveys/acs/summary_file/2018/data/1_year_ranking/}.
\newblock {Accessed: 26 July 2020}.

\bibitem{DemographicData2020201820162014201220082004}
{\sc {U.S. Census Bureau, Population Division}}, {\em {Annual Estimates of the
  Resident Population by Sex, Race, and Hispanic Origin: April 1, 2010 to July
  1, 2019}}.
\newblock
  \url{https://www.census.gov/data/tables/time-series/demo/popest/2010s-state-detail.html},
  2019.
\newblock {Last accessed: 25 July 2020}.

\bibitem{DemographicData2022}
{\sc {U.S. Census Bureau, Population Division}}, {\em {Annual Estimates of the
  Resident Population by Sex, Race, and Hispanic Origin: April 1, 2020 to July
  1, 2021, 2021 Column (as of July 1)}}.
\newblock
  \url{https://www.census.gov/data/tables/time-series/demo/popest/2020s-state-detail.html},
  2022.
\newblock {Last accessed: 24 July 2022}.

\bibitem{DemographicData2024}
{\sc {U.S. Census Bureau, Population Division}}, {\em {Annual Estimates of the
  Resident Population by Sex, Race, and Hispanic Origin: April 1, 2020 to July
  1, 2023}}.
\newblock
  \url{https://www.census.gov/data/tables/time-series/demo/popest/2020s-state-detail.html},
  2022.
\newblock {Last accessed: 26 August 2024}.

\bibitem{VolkeningChapter}
{\sc A.~Volkening}, {\em A primer on data-driven modeling of complex social
  systems}.
\newblock \url{https://arxiv.org/abs/2210.08636}, 2023.

\bibitem{Gitlab_elections}
{\sc A.~Volkening, D.~F. Linder, M.~A. Porter, and G.~A. Rempala}, {\em {Code
  associated with ``Forecasting Elections Using Compartmental Models of
  Infection"}}.
\newblock
  \url{https://gitlab.com/alexandriavolkening/forecasting-elections-using-compartmental-models},
  2020.

\bibitem{Volkening2020}
{\sc A.~Volkening, D.~F. Linder, M.~A. Porter, and G.~A. Rempala}, {\em
  Forecasting elections using compartmental models of infection}, SIAM Rev, 62
  (2020), pp.~837--865.

\bibitem{PrincetonElectionConsortium}
{\sc S.~Wang}, {\em {Princeton Election Consortium: Innovations in democracy
  since 2004}}.
\newblock \url{https://election.princeton.edu}, 2024.
\newblock Last accessed: 8 May 2024.

\bibitem{Wang}
{\sc S.~S.-H. Wang}, {\em Origins of {P}residential poll aggregation: {A}
  perspective from 2004 to 2012}, Int J Forecast, 31 (2015), pp.~898--909.

\bibitem{Westwood}
{\sc S.~J. Westwood, S.~Messing, and Y.~Lelkes}, {\em Projecting confidence:
  How the probabilistic horse race confuses and demobilizes the public}, J
  Politics, 82 (2020), pp.~1530--1544.

\bibitem{Wlezien1996}
{\sc C.~Wlezien and R.~S. Erikson}, {\em Temporal horizons and presidential
  election forecasts}, American politics quarterly, 24 (1996), pp.~492--505.

\bibitem{Wlezien}
{\sc C.~Wlezien and R.~S. Erikson}, {\em {The Timeline of Presidential Election
  Campaigns}}, J Politics, 64 (2002), pp.~969--993.

\bibitem{Wlezien2004}
{\sc C.~Wlezien and R.~S. Erikson}, {\em The fundamentals, the polls, and the
  presidential vote}, PS: Political Sci Politics, 37 (2004), pp.~747--751.

\bibitem{Yang2024}
{\sc F.~Yang, M.~Cai, C.~Mortenson, H.~Fakhari, A.~D. Lokmanoglu, J.~Hullman,
  S.~Franconeri, N.~Diakopoulos, E.~C. Nisbet, and M.~Kay}, {\em Swaying the
  public? impacts of election forecast visualizations on emotion, trust, and
  intention in the 2022 u.s. midterms}, IEEE Trans Vis Comput Graph, 30 (2024),
  pp.~23--33.

\bibitem{Yang2020}
{\sc V.~C. Yang, D.~M. Abrams, G.~Kernell, and A.~E. Motter}, {\em {Why Are
  U.S. Parties So Polarized? A ``Satisficing" Dynamical Model}}, SIAM Rev, 62
  (2020), pp.~646--657.

\end{thebibliography}

\end{document}